\documentclass{article} % For LaTeX2e
\PassOptionsToPackage{table}{xcolor}
\usepackage{iclr2027_conference,times}
\usepackage{amsmath,amsfonts,bm}

\def\eqref#1{equation~\ref{#1}}
\def\1{\bm{1}}

\DeclareMathAlphabet{\mathsfit}{\encodingdefault}{\sfdefault}{m}{sl}
\SetMathAlphabet{\mathsfit}{bold}{\encodingdefault}{\sfdefault}{bx}{n}

\usepackage{hyperref}
\usepackage{url}
\usepackage{booktabs}
\usepackage{amsmath}
\usepackage{amssymb}
\usepackage{graphicx}
\usepackage{xcolor}
\usepackage{colortbl}
\usepackage{tcolorbox}
\usepackage{wrapfig}
\usepackage{float}
\usepackage{makecell}
\usepackage{enumitem}
\usepackage{tikz}
\usepackage{subcaption}
\usepackage{xcolor} \usepackage{soul} 
\definecolor{ControlColor}{HTML}{9C93F8} 
\definecolor{ContentColor}{HTML}{5ECEBD} 
\newcommand{\controlhl}[1]{
\begingroup 
\sethlcolor{ControlColor} \hl{#1} \endgroup } \newcommand{\contenthl}[1]{
\begingroup 
\sethlcolor{ContentColor} \hl{#1} \endgroup }

\usetikzlibrary{positioning,arrows.meta,fit,backgrounds,patterns,decorations.pathreplacing,calc}

\definecolor{pastelblue}{HTML}{E7ECFE}

\title{HiPLEX: Hierarchical Policy Factorization for Full Duplex Speech Language Models
}

\author{
Kyudan Jung$^{1,2}$\thanks{This publication was created by Kyudan Jung while an intern at Qualcomm Technologies, Inc. (QTI) and currently attends Korea Advanced Institute of Sceience and Technology (KAIST).}, 
Hyunsin Park$^1$, 
Yoonhyung Lee$^1$,
Jinhwan Park$^1$,\\\textbf{Jinhyeok Yang$^1$,}
\textbf{KiHyun Nam$^1$,}
\textbf{Jaegul Choo$^2$,}
\textbf{Jinkyu Lee$^1$}\\
\textsuperscript{1}Qualcomm AI Research, \textsuperscript{2}KAIST AI\\
\small\texttt{kyudan@kaist.ac.kr}\\
\small\texttt{\{hyunsinp,jinhpark,yoonhyun,jinhyang,kihynam\}@qti.qualcomm.com,}\\
\small\texttt{jchoo@kaist.ac.kr, jinkyu@qti.qualcomm.com}
}

\newcommand{\method}{\textsc{HiPLEX}}

\newcommand{\pad}{\ensuremath{\mathtt{pad}}}
\newcommand{\epad}{\ensuremath{\mathtt{epad}}}
\newcommand{\con}{\ensuremath{\mathtt{cont}}}

\iclrfinalcopy
\begin{document}
\maketitle

\begin{abstract}
As human--AI interactions become more conversational, full-duplex speech language models capable of natural real-time dialogue are growing in importance. Beyond generating appropriate responses, these models must coordinate turn-taking, backchanneling, and floor management in real time.
Reinforcement learning (RL) provides a way to refine these behaviors through direct feedback on interaction outcomes. However, existing RL methods either apply timing feedback to a token policy or optimize semantic content, leaving the joint improvement of timing and content unresolved.
We introduce \method{}, an RL framework that factorizes a pretrained full-duplex text policy into a control policy that decides when to emit content and a conditional content policy that decides what to emit. The first factor selects among \pad{}, \epad{}, and \con{}. The second selects a token only when \con{} is chosen. This hierarchy describes conditional actions within each frame and uses the model's existing text head.
We route timing advantages to the token-group factor through event-causal masks derived from generated speech episodes, and route an LLM-judge semantic advantage to the conditional content factor. Across three Moshi seeds on Full-Duplex-Bench v1, \method{} reduces takeover rates during natural user pauses and backchannel opportunities, and shortens post-interruption response latency relative to GRPO, while maintaining comparable judged interruption-response quality. On Moshi and PersonaPlex, \method{} better matches pooled human turn-timing and backchannel-rate marginals than GRPO.
\end{abstract}

\section{Introduction}

Speech is an intrinsically duplex medium. Humans listen while preparing to speak, produce short acknowledgments without taking the floor, stop when interrupted, and resume when the floor becomes available~\citep{skantze2021review}. Recent full-duplex speech language models (SLMs) move toward this interaction pattern by generating audio in a streaming fashion while continuously conditioning on incoming user speech \citep{defossez2024moshi, roy2026personaplexvoicerolecontrol, chen2025minmomultimodallargelanguage}. However, learning such behavior remains difficult: the model must decide not only \emph{what} to say, but also \emph{when}, \emph{how long}, and \emph{under which conversational floor state} to say it.
%ok
Recent full-duplex RL methods already separate timing from content. ASPIRin projects the action space into binary active/inactive speech states \citep{hsiao2026aspirin}, DuplexPO optimizes purely timing-based rewards with GRPO inside fixed \emph{dynamics-critical windows} around annotated events, keeping semantics out of the RL objective \citep{li2026duplexpo}. Yet two structural questions remain. First, \textbf{where should credit land within an event?} A fixed window applies one advantage to every token inside it, yet the frames that cause a late onset differ from those that cause a well-timed one, blaming the wrong frames pushes timing the wrong way.
Second, \textbf{must content be excluded from RL to be protected?} If nothing in the reward asks for a better answer, content can stay at the level supervised fine-tuning left it, and it could even degrade: the timing reward still flows through the same policy that selects content tokens, and we show that timing-only training degrades judged quality while timing improves.

We treat full-duplex speech RL as \textbf{a hierarchical credit-assignment problem}, in the spirit of hierarchical advantage estimation for LLM agents \citep{peng2026hiper}, resolved with a mechanism specific to streaming speech. At each frame the high level decides whether and when content emission is invoked: the text stream emits padding tokens as placeholders when the model is not speaking (listening), exits padding (about to speak), or enters the text content partition (speaking). Only in the last case is the low-level content decision invoked. \method{} derives these two conditional decisions from a Moshi~\citep{defossez2024moshi}-style softmax, taking \pad/\epad/\con{} marginals via log-sum-exp and content tokens conditional inside the \con{} partition. Thus this is a conditional per-frame action hierarchy---\emph{whether/when to emit} above \emph{what to emit}---rather than an option hierarchy or a pair of separate network modules. Both factors come from the same pretrained head and introduce no new parameters. Each decomposed interaction reward (turn onset, pause, backchannel, yield) is then routed through an \emph{event-causal credit mask}, a sign-dependent eligibility rule computed from the policy's own generated speech episodes. An ASR-plus-LLM-judge semantic reward is routed to the conditional content factor. The result is a high-level objective for when to emit and a low-level objective for what to emit, rather than one advantage broadcast over the flat token stream.

We make three contributions:
\begin{itemize}
    \item \textbf{A two-level conditional action hierarchy with no architectural change.} We factorize the pretrained text policy into a high-level decision over whether and when to invoke content emission and, when \con{} is selected, a low-level decision over which content token to emit. This exposes a hierarchy already implicit in generation and requires no new parameters.
    \item \textbf{Credit that follows event structure, not proximity.} Our \emph{event-causal credit masks} are sign-dependent and read off the policy's own speech episodes: a late response penalizes the waiting decisions that delayed it, whereas a timely response credits its onset. A fixed window cannot make this distinction.
    \item \textbf{A frontier full-duplex model post-training method with comprehensive analysis.} On Full-Duplex-Bench v1, \method{} gives the strongest overall trade-off on Moshi and improves the restraint axes on PersonaPlex. We report a complementary post-boundary word-activity diagnostic for the mixed PersonaPlex turn results and analyze the interaction styles that different metrics favor.
\end{itemize}

\section{Related Work}
\vspace{-10pt}
\paragraph{RL for full-duplex interaction.}
Speech language models pursue end-to-end full duplex system to implement human-like behavior and gain low latency \citep{defossez2024moshi,chen2025minmomultimodallargelanguage, roy2026personaplexvoicerolecontrol, kim2026raonspeechtechnicalreport, wang2026covoaudiotechnicalreport}. Moshi interleaves an inner-monologue text stream with streaming audio generation and is the substrate most full-duplex work builds on \citep{defossez2024moshi}. PersonaPlex adds prompt-based voice and role control on top of it \citep{roy2026personaplexvoicerolecontrol}, and related systems target simultaneous listening and speaking, backchannels, and interruption handling \citep{ma2024lslm,veluri2024syncllm}. We assume this modeling and study the RL objective that refines its interaction dynamics. Raw-token RL can improve turn timing while degrading content through repetition and drift~\citep{hsiao2026aspirin}. Kyutai team applies GRPO to Moshi with four faceted rewards~\citep{ohashi2026multifacetedinteractivityalignmentfullduplex}. DuplexPO applies GRPO with timing-only factorized rewards inside fixed dynamics-critical windows around annotated events~\citep{li2026duplexpo}.
\vspace{-10pt}
\paragraph{Hierarchical RL and credit assignment.}
Hierarchical reinforcement learning introduces structured decisions through options, subpolicies, or manager--worker decompositions \citep{sutton1999between,bacon2017option,kulkarni2016hierarchical}. For language agents, flat policy-gradient methods must propagate sparse rewards across long trajectories, which can be unstable~\citep{peng2026hiper}. HiPER aligns credit assignment with high-level subgoal segments and low-level actions via Hierarchical Advantage Estimation \citep{peng2026hiper}. \method{} adapts the principle that credit should follow the decision structure rather than the token sequence to streaming speech. Its hierarchy is conditional rather than option-based: the high-level \pad/\epad/\con{} action determines whether a low-level content decision is invoked at that frame. The token-group factorization uses the pretrained text head without an added manager, and audio-grounded event masks replace learned segment values. Appendix~\ref{app:related} gives extended related works.

\begin{figure*}
    \centering
    %\vspace{-10pt}
    \includegraphics[width=1\linewidth]{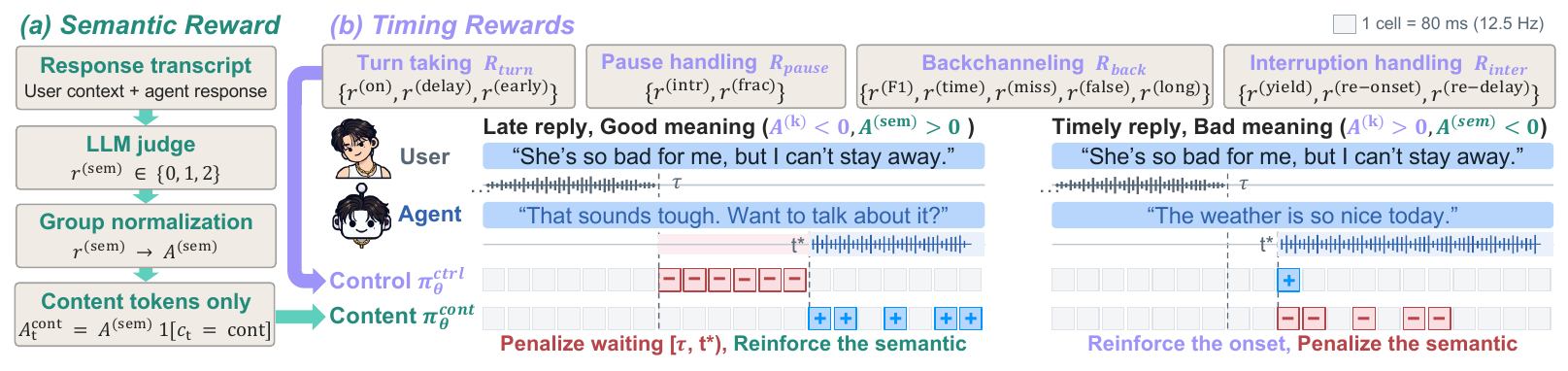}
    \vspace{-20pt}
    \caption{Timing and semantic credit assignment in HiPLEX. (a) An LLM judge evaluates response relevance and coherence. HiPLEX normalizes these scores within each rollout group and applies the semantic advantage to the conditional content policy only where \(c_t=\mathrm{cont}\). (b) Timing components for the active interaction type are normalized separately, and event-causal masks route their advantages to control decisions. Each cell represents an 80-ms frame.}
    \label{fig:overview}
\end{figure*}

\vspace{-10pt}
\section{Method}
\vspace{-10pt}
\method{} trains a full-duplex speech language model by assigning timing feedback to a control policy and semantic feedback to a conditional content policy.
We first define the streaming setting and derive both policies from the model's existing text distribution. We then describe the rewards, explain how they are assigned to individual decisions, and present the training objective. 

\vspace{-10pt}
\subsection{Problem Setup}

We consider a full-duplex speech language model that processes and generates audio in $80$\,ms frames. At frame $t$, the model is conditioned on user audio $x_{\leq t}$, its own past audio $a_{<t}$, its past text tokens $w_{<t}$, and optional role or voice conditioning $p$~\citep{roy2026personaplexvoicerolecontrol}. These inputs form the causal state $s_t=(x_{\leq t},a_{<t},w_{<t},p)$.
In Moshi-style models, the text stream follows the same clock as the audio stream. We refer to text tokens other than \pad{} and \epad{} as \emph{content tokens}. The \pad{} token fills a frame without a new content token, and \epad{} marks the end of a padding span. Padding can occur both during silence and within ongoing speech. Thus an individual text token does not reliably indicate whether the model is currently producing speech.

We train on annotated interaction windows covering four
conversational behaviors: turn taking, pause handling, backchanneling, and interruption handling.
These windows specify when the model should acknowledge
the user, respond after a user turn, remain silent during
a user hesitation, or yield to an interruption and
re-enter after it ends.
Appendix~\ref{app:data} details how these windows are
extracted from human conversations.

We measure the model's timing behavior from its generated
audio. We extract speech spans with Silero VAD~\citep{silero2024vad} and merge consecutive
spans separated by gaps of at most $1$\,s into
\emph{speech episodes}. Episodes lasting at most $1$\,s
are classified as \emph{short}, and longer episodes as
\emph{sustained}. Short episodes are used to identify
candidate backchannels, while sustained episodes are
used to identify turn and re-entry onsets.
Together with the annotated event times and intervals,
these episodes provide a common temporal basis for
computing timing rewards across all four behaviors and
selecting the control decisions that receive credit
(Section~\ref{sec:eventcausal}).

\subsection{Control and Content Policy Factorization}
\label{sec:factorization}

\begin{wrapfigure}{r}{0.38\linewidth}
\vspace{-18pt}
    \centering
    \includegraphics[width=\linewidth]{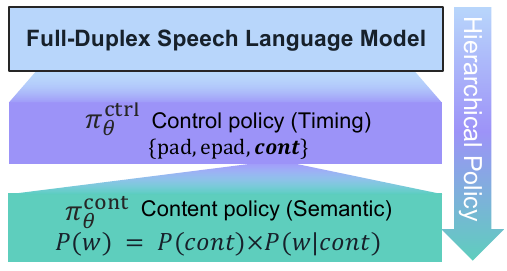}
    \vspace{-20pt}
    \caption{Hierarchical policy factorization in HiPLEX. }
    \vspace{-20pt}
    \label{fig:hierarchical}
\end{wrapfigure}

We factorize the text policy into a high-level \controlhl{\textit{control policy}} and a low-level \emph{\contenthl{content policy}} \con{} as shown in Figure~\ref{fig:hierarchical}. The control policy selects a token group at each frame. The content policy selects a token only when the control policy selects the content group. We partition the vocabulary into the singleton groups $\mathcal{V}_{\pad}=\{\pad\}$ and $\mathcal{V}_{\epad}=\{\epad\}$, and the content group $\mathcal{V}_{\con}$.

Let $\ell_t(v)$ denote the logit used to form text-token probabilities at frame $t$. In our training runs, we divide the model's raw text logits by the text sampling temperature $T_{\mathrm{text}}=0.7$ before constructing either factor. The control policy assigns each group its total probability under the resulting flat text distribution. We compute this group probability using a log-sum-exp over its tokens and a softmax over the three groups.
\begin{equation}
    g_t(c)=\log\!\!\sum_{v\in\mathcal{V}_c}\!\exp \ell_t(v),\qquad
    \pi_{\theta}^{\mathrm{ctrl}}(c_t\!=\!c\mid s_t)=\mathrm{softmax}(g_t)_c,\quad c\in\{\pad,\epad,\con\}.
\end{equation}
When $c_t=\con$, the content policy is the conditional distribution within $\mathcal{V}_{\con}$.
\begin{equation}
    \pi_{\theta}^{\con}(w_t\!=\!v\mid s_t,c_t\!=\!\con)=\frac{\exp \ell_t(v)}{\sum_{u\in\mathcal{V}_{\con}}\exp \ell_t(u)},\qquad v\in\mathcal{V}_{\con}.
    \label{eq:factorization}
\end{equation}

For a content token $v$, the two factors multiply to recover its probability under this text distribution.
\begin{equation}
    \pi_\theta(w_t=v\mid s_t)
    =\pi_\theta^{\mathrm{ctrl}}(c_t=\con\mid s_t)
    \pi_\theta^{\con}(w_t=v\mid s_t,c_t=\con).
\end{equation}
For \pad{} and \epad{}, the control probability alone equals the corresponding token probability. This factorization reconstructs the temperature-scaled text-token distribution and introduces no new parameters. Appendix~\ref{app:rldiag} reports a numerical reconstruction check.

This hierarchy describes whether to emit a content token and which token to emit at each frame. The speech episodes connect these token decisions to conversational timing. We assign timing advantages to the control log-probability and semantic advantages to the conditional content log-probability. The control and content policies are two factors of a
single model's text distribution and share the same parameters. Thus, although timing and semantic rewards are assigned to different policy factors, optimizing either factor can also affect the other.

\subsection{Rewards and Credit Assignment}
\label{sec:eventcausal}
\vspace{-2pt}
For each annotated interaction window, we sample a group of $G$ rollouts from the current sampling policy $\pi_{\theta_{\mathrm{old}}}$. The rollouts share the same user context and annotated event. We score their timing with reward components and their response quality with a semantic reward. Each score is normalized within this group before it is assigned to policy decisions. Figure~\ref{fig:overview} illustrates this procedure. 

\paragraph{Timing rewards.}
We use three timing-reward components for turn taking, two for pause handling, five for backchanneling, and three for interruption handling. Each component assigns a scalar score to a rollout by comparing the generated speech with the annotated event.
Only the components corresponding to the sampled window's interaction type are used. To illustrate the reward design, we describe the turn-taking components in detail and briefly summarize the remaining three interaction types. Appendix~\ref{app:rewards} provides the precise definitions, matching rules, weights, and edge cases for all components. Appendix~\ref{app:hparams} lists the training hyperparameters.

\begin{itemize}
% [leftmargin=0pt,itemindent=0pt,labelsep=0.5em,itemsep=1pt]
\item \textbf{Turn taking} ($R_{turn}=\{r^{(\mathrm{on})}, r^{(\mathrm{delay})}, r^{(\mathrm{early})}\}$), the three components score response onset ($r^{(\mathrm{on})}$), response delay ($r^{(\mathrm{delay})}$), and early speech ($r^{(\mathrm{early})}$).
Let $\tau$ be the end of the user's turn and $T$ the end of the interaction window, measured from the window start.
Let $t^\star$ be the earliest onset in $[\tau,T]$ of a speech episode lasting longer than $1$\,s, if one exists.
The onset and delay scores are
  \begin{equation}
  \begin{aligned}
  r^{(\mathrm{on})}
  &=
  \begin{cases}
  \exp\left[-\dfrac{(t^\star-\tau)^2}{2\sigma^2}\right]
  +\beta\exp\left[-\dfrac{t^\star-\tau}{\kappa}\right],
  & \text{if a qualifying episode exists},\\
  -0.5,
  & \text{otherwise},
  \end{cases}\\[-1pt]
  r^{(\mathrm{delay})}
  &=
  \begin{cases}
  -\min\left(1,\dfrac{t^\star-\tau}{T-\tau}\right),
  & \text{if a qualifying episode exists},\\
  -1,
  & \text{otherwise}.
  \end{cases}
  \end{aligned}
  \label{eq:onset}
  \end{equation}
The onset component $r^{(\mathrm{on})}$ favors responses close to $\tau$ and gives any valid response a higher score than a missing response. The parameters $\sigma$ and $\kappa$ control how quickly the two terms decay, and $\beta$ weights the second term. The delay component $r^{(\mathrm{delay})}$ decreases linearly with response delay relative to the remaining window length. We standardize their nonlinear onset and linear delay signals separately before weighting. To discourage premature responses, we assign an early-start penalty $r^{(\mathrm{early})}=-0.5\rho_{\mathrm{early}}$, where $\rho_{\mathrm{early}}$ is the duration of model speech in $[0,\tau)$ divided by $\tau$. Figure~\ref{fig:timing-score-shapes} illustrates these scores.
The remaining interaction types use the following components.

\item \textbf{Pause } ($R_{pause}=\{r^{(\mathrm{intr})},r^{(\mathrm{frac})}\} $). A binary intrusion score penalizes any speech episode lasting at least
$1$\,s during a user hesitation. A speech-fraction penalty also discourages shorter intrusions by measuring
the fraction of the window occupied by model speech.
\item \textbf{Backchannel} ($R_{back}=\{r^{(\mathrm{F1})}, r^{(\mathrm{time})}, r^{(\mathrm{miss})}, r^{(\mathrm{false})}, r^{(\mathrm{long})}\}$). An F1 score measures how well short episodes match annotated
backchannel opportunities, following \citet{ohashi2026multifacetedinteractivityalignmentfullduplex}.
A separate timing score decreases with the distance from each opportunity to the nearest short episode.
Three penalties cover missed opportunities, unmatched short episodes (false alarms), and episodes longer than
$1$\,s (overlong acknowledgments).
\item \textbf{Interruption} ($R_{inter}=\{r^{(\mathrm{yield})}, r^{(\text{re-onset})}, r^{(\text{re-delay})}\}$). A yield score penalizes model speech that overlaps the user's
interruption after a $0.16$\,s grace period. Re-entry onset and delay use Eq.~\ref{eq:onset}, with $\tau$
set to the end. They score the first sustained response that starts at or after the anchor.
\end{itemize}

\paragraph{Group-relative advantages.}
We use parenthesized superscripts to identify reward components and subscripts to index rollouts. For example, $r_i^{(\mathrm{on})}$ denotes the onset reward for rollout $i$. We omit the rollout index when defining reward components.
For each active timing component $k$ defined above, let $r_i^{(k)}$ denote its score for rollout $i$
among the $G$ rollouts sampled for the same interaction window.
Following~\citet{shao2024deepseekmath}, we standardize each component independently within this group.
Let $\bar{r}^{(k)}$ and $s_r^{(k)}$ denote the mean and population standard deviation of
$\{r_j^{(k)}\}_{j=1}^{G}$. The component advantage is
\begin{equation}
    A_i^{(k)}=\frac{r_i^{(k)}-\bar{r}^{(k)}}{\max(s_r^{(k)},\delta)},
    \qquad i=1,\ldots,G.
    \label{eq:group-advantage}
\end{equation}
The constant $\delta=10^{-8}$ stabilizes the denominator.
Independent normalization makes components with different reward scales comparable.
If all rollouts receive the same score for a component, its advantage is zero.
For designated sparse components, we replace this zero advantage with a fixed negative value when all
rollouts fail. Appendix~\ref{app:rewards} specifies the affected components, activation conditions, and constants.

\paragraph{Event-causal credit masks.}
The group-relative advantage provides one score per reward component for each rollout.
We next determine which control decisions within that rollout should receive this score.
We select these decisions using the model's speech episodes, the ground-truth event timing, and the sign of the component advantage.
A binary mask $M_t^{(k)}$ is one at selected frames and zero elsewhere.
Omitting the rollout index $i$, the advantage applied to the control policy $\pi_\theta^{\mathrm{ctrl}}$ at frame $t$ is
\begin{equation}
    A_t^{\mathrm{ctrl}}=\sum_{k}w_k A^{(k)}\,M^{(k)}_t.
    \label{eq:routing}
\end{equation}
Here $w_k$ weights component $k$. The selected frames follow the speech behavior relevant to that component,
such as waiting before a response or continuing to speak during an interruption.
We call these \emph{event-causal masks} because they assign credit according to the role of each control decision
in the observed timing outcome. Speech episodes distinguish waiting from padding within ongoing speech.
All episode boundaries, advantages, and masks are computed after rollout and held fixed during the policy update.
%
% For an ordinarily normalized component, a positive advantage means that its score is above the rollout-group mean.
The masking rules for each interaction axis are as follows:
\begin{itemize}
% [leftmargin=0pt,itemindent=0pt,labelsep=0.5em,itemsep=1pt]
\item \textbf{Turn taking.} Positive onset or delay advantages reinforce the decision to begin speaking.
Non-positive advantages select the preceding waiting interval, so a late-response penalty discourages waiting
without suppressing the eventual response. Early-start penalties target speech before the user's turn ends.
\item \textbf{Pause.} Intrusion penalties target the speech produced in the pause-handling window.
For a silent rollout, positive advantages reinforce the control decisions within the ground-truth pause windows.
\item \textbf{Backchannel.} Positive timing credit targets a matching short episode.
Miss penalties target opportunity windows, false-alarm penalties target unmatched episodes, and overlong
penalties target speech beyond the allowed duration.
\item \textbf{Interruption.} Failure-to-yield penalties target speech overlapping the user.
For a rollout that yields, positive advantages reinforce silence during the interruption.
Re-entry uses the same onset-versus-waiting rule as turn taking.
\end{itemize}
\vspace{-5pt}
If several components select the same frame, their weighted advantages add.
Figure~\ref{fig:overview} illustrates these credit locations.
Appendix~\ref{app:masks} specifies the component-level masks, temporal boundaries, and missing-event cases.

%\vspace{-20pt}
\paragraph{Semantic reward.}
\label{sec:semantic}
We assess response relevance and coherence by transcribing the user context and generated speech with
Parakeet-TDT-0.6B-v2~\citep{nvidia2025parakeet} and asking Gemini~2.5 Flash-Lite~\citep{google2025geminiflashlite}
to assign a response-level score $r_i^{(\mathrm{sem})}\in\{0,1,2\}$.
We standardize these scores across the $G$ rollouts for the same context to obtain
$A_i^{(\mathrm{sem})}=(r_i^{(\mathrm{sem})}-\bar r^{(\mathrm{sem})})/\max(s_r^{(\mathrm{sem})},\delta)$,
where $\bar r^{(\mathrm{sem})}$ and $s_r^{(\mathrm{sem})}$ are the group mean and population standard deviation,
and $\delta$ is the same denominator floor used for timing advantages.
Although the judge scores the response as a whole, we apply its advantage only to the conditional content
policy $\pi_\theta^{\con}$ at frames where a content token was generated. Omitting the rollout index, this gives
$A_t^{\con}=A^{(\mathrm{sem})}\,\mathbf{1}[c_t=\con]$.
Appendix~\ref{app:judgeprotocol} gives the scoring prompt and protocol.

\subsection{Objective and Learning Rate}
\label{sec:objective}

We optimize $\pi_{\theta}^{\mathrm{ctrl}}$ and $\pi_{\theta}^{\mathrm{cont}}$ using their respective advantages. Each policy has its own importance ratio relative to the policy that generated the rollouts.
\begin{equation}
    \rho_t^{\mathrm{ctrl}}=\frac{\pi_{\theta}^{\mathrm{ctrl}}(c_t\mid s_t)}{\pi_{\theta_{\mathrm{old}}}^{\mathrm{ctrl}}(c_t\mid s_t)},\qquad
    \rho_t^{\con}=\frac{\pi_{\theta}^{\con}(w_t\mid s_t,c_t\!=\!\con)}{\pi_{\theta_{\mathrm{old}}}^{\con}(w_t\mid s_t,c_t\!=\!\con)}.
\end{equation}
The content ratio is evaluated only when $c_t=\con$. For factor $z\in\{\mathrm{ctrl},\con\}$, let $m_t^z$ be a binary mask indicating whether the decision at frame $t$ is included in that factor's policy loss. A value of one includes the decision, and zero excludes it. The control mask selects decisions covered by the event-causal masks, and the content mask selects content decisions. We use PPO-style clipped losses~\citep{schulman2017ppo}, normalized by the number of decisions selected for each factor.
\begin{equation}
\begin{aligned}
    \mathcal{L}_z&=-\frac{1}{\sum_t m_t^z}\sum_t m_t^z
    \min\!\left(\rho_t^z A_t^z,\ \mathrm{clip}(\rho_t^z,1\!-\!\epsilon,1\!+\!\epsilon)A_t^z\right),\\
    \mathcal{L}&=\alpha_c\mathcal{L}_{\mathrm{ctrl}}+\alpha_l\mathcal{L}_{\con}+\mathcal{L}_{\mathrm{KL}}.
\end{aligned}
\end{equation}
Here $\epsilon$ is the clipping range, and $\alpha_c$ and $\alpha_l$ weight the two losses. A factor contributes no policy loss when it has no selected decisions. Separate normalization keeps the content loss from being diluted by long padding spans. We also regularize each policy toward a frozen reference model, following the reference-policy regularization in~\citet{shao2024deepseekmath}:
\begin{equation}
\begin{aligned}
\mathcal{L}_{\mathrm{KL}}
&=\beta_c\mathrm{KL}(\pi_\theta^{\mathrm{ctrl}}\|\pi_{\mathrm{ref}}^{\mathrm{ctrl}})+\beta_l\pi_\theta^{\mathrm{ctrl}}(\con)\mathrm{KL}(\pi_\theta^{\con}\|\pi_{\mathrm{ref}}^{\con}).
\end{aligned}
\end{equation}
The coefficients $\beta_c$ and $\beta_l$ control the strength of regularization for the two policies. Training updates all model parameters in a single phase. Appendix~\ref{app:algorithm} gives the full training loop.

% Appendix~\ref{app:rldiag} discusses how this factor-wise objective differs from flat GRPO despite the policy factorization.

\paragraph{Learning rate.}
Event-causal masking selects approximately $20$--$30\%$ of the frames used by a flat update in our reported runs. Normalizing each loss by its selected decisions controls the loss scale, but the two objectives still update different decisions at different frequencies. We therefore tune their learning rates separately. For Moshi, validation selects $4\times10^{-6}$ for \method{} and $1\times10^{-6}$ for the flat GRPO baseline. PersonaPlex uses $2\times10^{-6}$ for \method{} with the reward settings in Appendix~\ref{app:hparams}. Appendix~\ref{app:lrsweep} reports the Moshi baseline sweep over the same learning-rate range.

\begin{table*}[t]
\centering
\small
\resizebox{\linewidth}{!}{%
\begin{tabular}{l cc ccc cc ccc}
\toprule
 & \multicolumn{2}{c}{\textbf{Pause}} & \multicolumn{3}{c}{\textbf{Backchannel}} & \multicolumn{2}{c}{\textbf{Smooth Turn Taking}} & \multicolumn{3}{c}{\textbf{User Interruption}} \\
\cmidrule(lr){2-3}\cmidrule(lr){4-6}\cmidrule(lr){7-8}\cmidrule(lr){9-11}
\textbf{Model} & Syn TOR$^{*}\downarrow$ & Candor TOR$\downarrow$ & TOR$\downarrow$ & Freq$\uparrow$ & JSD$\downarrow$ & TOR$\uparrow$ & Latency$\downarrow$ & TOR$\uparrow$ & Judge$\uparrow$ & Latency$\downarrow$ \\
\midrule
Moshi~\citep{defossez2024moshi} & \textbf{0.416} & 0.579 & 0.473 & \underline{0.126} & \textbf{0.724} & \underline{0.387} & \textbf{0.000} & 0.772 & 3.773 & 0.852 \\
+ GRPO & \underline{1.000} & \underline{0.454} & \underline{0.309} & 0.110 & \underline{0.737} & \textbf{0.992} & \textbf{0.000} & \underline{0.955} & \underline{3.942} & \underline{0.660} \\
\rowcolor{pastelblue}+ \method{} & \underline{1.000} & \textbf{0.306} & \textbf{0.127} & \textbf{0.134} & 0.734 & \textbf{1.000} & \textbf{0.000} & \textbf{0.960} & \textbf{4.083} & \textbf{0.439} \\
\midrule
PersonaPlex~\citep{roy2026personaplexvoicerolecontrol} & \underline{0.847} & \underline{0.426} & \underline{0.218} & 0.103 & \underline{0.755} & \textbf{0.882} & \textbf{0.000} & 0.945 & 2.693 & 0.712 \\
+ GRPO & 0.876 & 0.444 & 0.236 & \textbf{0.107} & \textbf{0.746} & \textbf{0.882} & \textbf{0.000} & \underline{0.956} & \underline{2.712} & \textbf{0.548} \\
\rowcolor{pastelblue}+ \method{} & \textbf{0.715} & \textbf{0.218} & \textbf{0.018} & \underline{0.106} & 0.769 & \underline{0.714} & \underline{0.002} & \textbf{0.975} & \textbf{2.938} & \underline{0.590} \\
\midrule
\multicolumn{11}{c}{\emph{Reference numbers for the same RL recipe \citep{ohashi2026multifacetedinteractivityalignmentfullduplex}}} \\
Moshi + RL, as reported & 0.307 & 0.463 & 0.145 & 0.101 & 0.794 & 0.958 & 0.160 & 1.000 & 3.630 & 0.409 \\
Moshi + RL, released checkpoint & 1.000 & 0.597 & 0.255 & 0.157 & 0.711 & 0.992 & 0.000 & 0.955 & 3.984 & 0.530 \\
PersonaPlex + RL, as reported & 0.350 & 0.356 & 0.073 & 0.112 & 0.786 & 0.975 & 0.086 & 0.995 & 4.533 & 0.223 \\
PersonaPlex + RL, released checkpoint & 0.818 & 0.384 & 0.145 & 0.178 & 0.730 & 0.849 & 0.001 & 0.995 & 3.427 & 0.534 \\
\bottomrule
\end{tabular}%
}
\caption{Result of Full-Duplex-Bench v1. The trained Moshi rows are seed-42
terminal checkpoints. Three-seed results are reported in Table~\ref{tab:seed-replication}. The lower block contains calibration references and is not used as a training baseline. Best per metric within a family in bold, second best
underlined.}
\label{tab:fdb1_highlight}
\vspace{-10pt}
\end{table*}

\section{Experimental Setup}
\vspace{-2.5pt}
\paragraph{Models and data.}
We use Moshi \citep{defossez2024moshi} which generates speech and an aligned inner-monologue text stream on a $12.5$\,Hz frame clock and PersonaPlex \citep{roy2026personaplexvoicerolecontrol} which adds role-prompt and voice conditioning on the same backbone, which we hold fixed across training and evaluation so that any difference comes from the RL objective. All RL training uses conversation-only subsets of the Seamless-Interaction corpus \citep{seamless2025}, following \citet{ohashi2026multifacetedinteractivityalignmentfullduplex}. We extract $73{,}561$ annotated interaction windows, approximately balanced across the four conversational behaviors: turn taking, pause handling, backchanneling, and interruption handling. Appendix~\ref{app:data} gives the filtering, window rules, and statistics.

\paragraph{Baselines.}
For each model family, we compare \method{} with the pretrained base model and our reproduction of the GRPO-based training method of \citet{ohashi2026multifacetedinteractivityalignmentfullduplex} (Table~\ref{tab:fdb1_highlight}). This reproduced GRPO baseline uses the same pretrained backbone as \method{} and serves as our primary comparison. We also report Ohashi et al.'s published results alongside our evaluations of their released Seamless-trained checkpoints in the lower block of Table~\ref{tab:fdb1_highlight}. Appendix~\ref{app:lrsweep} examines the sensitivity of the reproduced Moshi GRPO baseline to its learning rate.
\vspace{-2.5pt}
\paragraph{Ablations.}
On Moshi, we first compare GRPO and \method{} with and without the semantic reward. We then hold the factorized policy and rewards fixed and compare four timing-credit rules: all-frame, fixed-window, random masking with matched sparsity, and event-causal masking. To examine the fixed-window credit assignment used in DuplexPO~\citep{li2026duplexpo}, we include a fixed-window variant under our training setup. We further examine the training-judge scale on Moshi and stronger semantic feedback on PersonaPlex (Appendices~\ref{app:rubric} and \ref{app:semmax}).
\vspace{-2.5pt}
\paragraph{Evaluation.} We evaluate on \textbf{Full-Duplex-Bench v1} \citep{lin2025fdbench}, which measures four full-duplex behaviors: holding silence during hesitation, backchanneling without taking the floor, taking turns after user yield, and responding to user interruptions. We follow the official streaming protocol with real-time playback from $t{=}0$ and no silence modification. Interruption latency measures the delay from the end of the user's interruption to the start of the model response, clamped at zero. Response quality (Judge) is scored by gpt-4o-2024-08-06~\citep{openai2024gpt4o} on frozen ASR transcripts (Appendix~\ref{app:judgeprotocol}). Since no single metric fully captures all interaction behaviors, we report results across multiple evaluation axes. For comparability, we report the official metrics, while details of the evaluation protocol and metric interpretation are provided in Appendices~\ref{app:pause} and \ref{app:turn}.

 \section{Results}
 %\vspace{-10pt}
\subsection{Main Comparison}
\begin{wrapfigure}{r}{0.45\linewidth}
\vspace{-62pt}
    \centering
    \includegraphics[width=\linewidth]{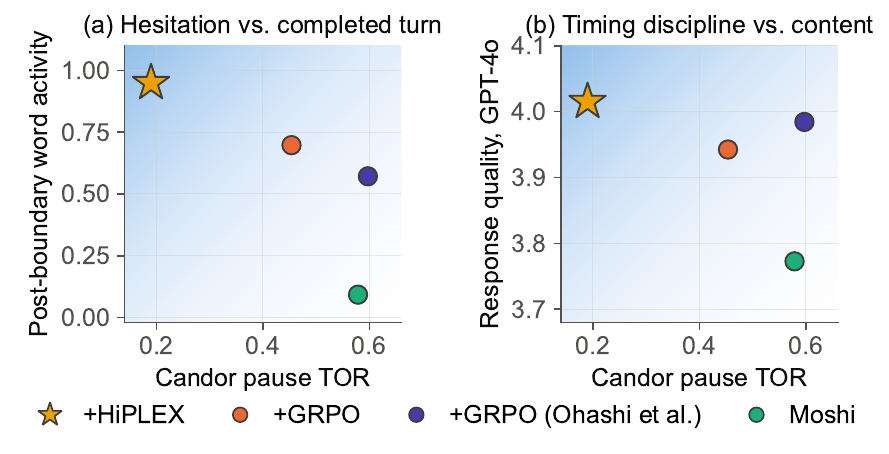}
    \vspace{-20pt}
    \caption{Moshi family on Full-Duplex-Bench v1. (a) shows post-boundary word activity, which may include speech initiated before the annotated turn end. (b) shows judged quality conditional on a qualifying interruption response. ``+GRPO (Ohashi et al.)'' is our evaluation of their released checkpoint.}
    \vspace{-15pt}
    \label{fig:frontier}
\end{wrapfigure}
\label{sec:results_main}
Table~\ref{tab:fdb1_highlight} reports Full-Duplex-Bench v1 results. The trained Moshi rows correspond to 100-epoch checkpoints, following the training duration of \citet{ohashi2026multifacetedinteractivityalignmentfullduplex}. On Moshi, \method{} reduces CANDOR pause and backchannel takeover rates and post-interruption response latency relative to GRPO. These gains hold across three training seeds, while mean judged response quality remains comparable (Table~\ref{tab:seed-replication}). The official smooth-turn TOR is nearly saturated for both methods and therefore provides limited evidence about turn-boundary behavior.
On PersonaPlex, \method{} also reduces pause and backchannel takeovers relative to GRPO. These restraint gains coexist with a lower official smooth-turn TOR and a longer post-interruption response latency. A complementary diagnostic (Appendix~\ref{app:turn}) shows more qualifying post-boundary word activity and a shorter median offset to the first post-boundary word for \method{} in both model families. On PersonaPlex, \method{} also begins speaking before the boundary less often. Together, these diagnostics show a more favorable balance of pre-boundary restraint and post-boundary word activity than the official TOR alone suggests.
\vspace{-5pt}
\paragraph{Pause restraint and response behavior.}
A user hesitation and a completed turn can both contain a brief silence, making them a joint test of when the model speaks. Figure~\ref{fig:frontier}(a) plots CANDOR pause TOR against post-boundary word activity rather than the nearly saturated official smooth-turn TOR. \method{} combines stronger pause restraint with more post-boundary word activity than GRPO. As discussed in Appendix~\ref{app:turn}, this activity can include speech initiated before the user turn ended and does not establish a higher rate of newly initiated responses. Figure~\ref{fig:frontier}(b) plots CANDOR pause TOR against the interruption-task Judge score, which is conditional on a qualifying model response. For the displayed checkpoint, \method{} combines stronger pause restraint with higher judged quality than GRPO. Across training seeds, mean judged quality remains comparable.

\subsection{The Same Recipe Needs a Different Step Size on the Two Families}
\label{sec:dynamics}
\vspace{-10pt}
\begin{figure}[H]
    \centering
    \includegraphics[width=\linewidth]{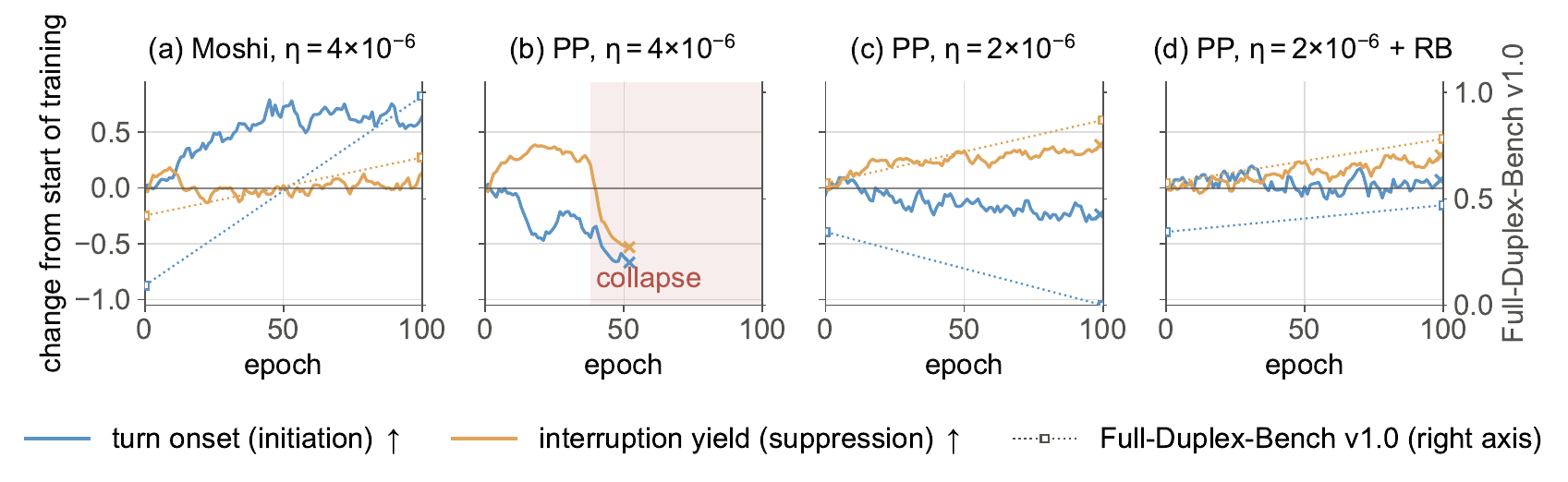}
    \vspace{-10pt}
    \caption{Reward during training with otherwise identical settings within each model family. Moshi is stable at $4\!\times\!10^{-6}$. PersonaPlex (PP) collapses, remains suppression-dominated at half that rate, and recovers with the rebalanced configuration (RB). Dotted lines show endpoint benchmark scores.}
    \label{fig:dynamics}
    \vspace{-15pt}
\end{figure}
\begin{wrapfigure}{r}{0.5\linewidth}
\vspace{-15pt}
    \centering
    \includegraphics[width=\linewidth]{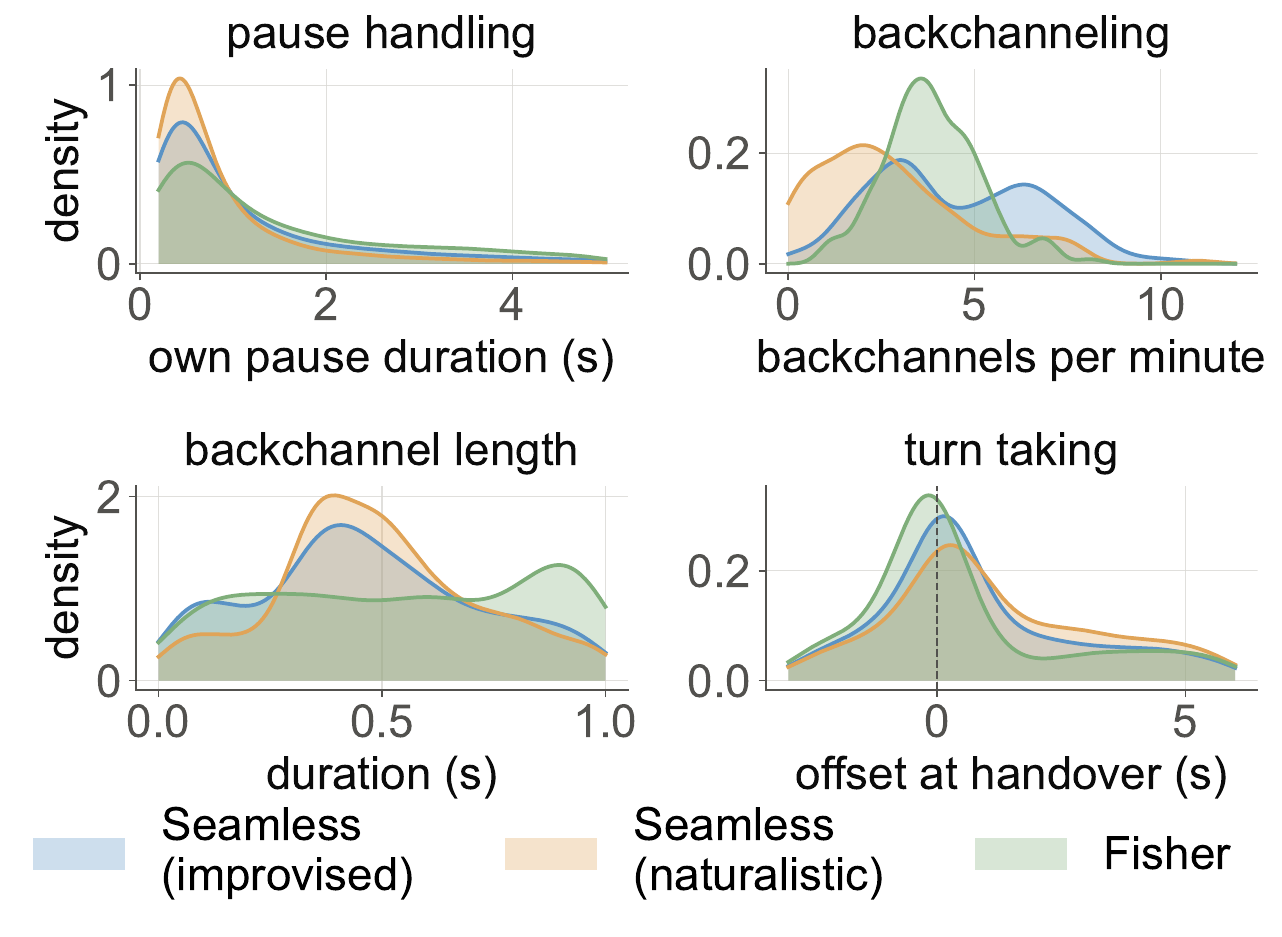}
    \vspace{-20pt}
    \caption{Human timing in the training corpora. Each panel is one quantity the benchmark scores.}
    \label{fig:datastats}
    \vspace{-10pt}
\end{wrapfigure}

Despite sharing Moshi's backbone, PersonaPlex is considerably more sensitive to the learning rate under the same training recipe, as shown in Figure~\ref{fig:dynamics}. At Moshi's learning rate of $4\times10^{-6}$, training collapses with suppression increasing while initiation steadily declines. Reducing the learning rate by half only delays convergence to the same near-silent behavior. We attribute this narrower optimization margin to PersonaPlex's more talkative, floor-holding prompt prior, which causes the early suppression signal to dominate learning. Rebalancing the rewards toward initiation prevents this collapse and keeps both rewards healthy throughout training. Additional diagnostic analyses are provided in Appendix~\ref{app:diagnostics}.

\subsection{Matching Human Timing, Not Just Passing a Rule}
\label{sec:distmatch}

Official scores largely ask whether an event occurred, whereas human timing is distributed. Figure~\ref{fig:datastats} shows broad corpus distributions for pauses, backchannel rate and length, and floor-transfer offset (Appendix~\ref{app:datastats} explains more detail about data statistics). Thus a model can pass a binary rule yet answer consistently late or acknowledge too often. Given this nature, we report Wasserstein-1 distance between model and pooled Seamless--Fisher marginals as shown in Table~\ref{tab:distmatch}.

\newpage
\vspace{-8pt}
\begin{wraptable}{r}{0.5\linewidth}
\centering
\centering
\small
\resizebox{\linewidth}{!}{%
\begin{tabular}{lcccc}
\toprule
Model & \makecell{Turn\\taking $\downarrow$} & \makecell{Backchannel\\rate $\downarrow$} & \makecell{Backchannel\\length $\downarrow$} & \makecell{Pause\\intrusion $\downarrow$} \\
\midrule
Moshi & 4.698 & 1.798 & 0.292 & 0.068 \\
\quad + GRPO & 1.781 & 1.512 & 0.266 & \textbf{0.012} \\
\rowcolor{pastelblue}\quad + \method{} & \textbf{0.902} & \textbf{0.327} & \textbf{0.238} & 0.080 \\
\midrule
PersonaPlex & 3.551 & 1.738 & 0.080 & \textbf{0.088} \\
\quad + GRPO & 3.511 & 1.652 & \textbf{0.054} & 0.109 \\
\rowcolor{pastelblue}\quad + \method{} & \textbf{2.627} & \textbf{1.187} & 0.168 & 0.091 \\
\bottomrule
\end{tabular}%
}
\vspace{-8pt}
\caption{Wasserstein-1 distance to pooled human timing marginals (lower is closer).}
\vspace{-12pt}
\label{tab:distmatch}

\end{wraptable}

 On Moshi, \method{} is closest for turn timing, backchannel rate, and backchannel length. GRPO is closest on pause intrusion because it is nearly always silent during hesitations, whereas the human rate is small but nonzero. PersonaPlex remains mixed: \method{} is closer on turn timing and backchannel rate, but not on the other two marginals. Unanswered turns receive a 6-s ceiling so silence cannot win by omission.

\subsection{Effect of Semantic Feedback}

\label{sec:ablations}
\begin{table}[h]
\vspace{-10pt}
\centering
\small
\resizebox{\linewidth}{!}{%
\begin{tabular}{l cc ccc cc ccc}
\toprule
 & \multicolumn{2}{c}{Pause} & \multicolumn{3}{c}{Backchannel} & \multicolumn{2}{c}{Smooth Turn Taking} & \multicolumn{3}{c}{User Interruption} \\
\cmidrule(lr){2-3}\cmidrule(lr){4-6}\cmidrule(lr){7-8}\cmidrule(lr){9-11}
Model & Syn TOR$\downarrow$ & Candor TOR$\downarrow$ & TOR$\downarrow$ & Freq$\uparrow$ & JSD$\downarrow$ & TOR$\uparrow$ & Latency$\downarrow$ & TOR$\uparrow$ & Judge$\uparrow$ & Latency$\downarrow$ \\
\midrule
GRPO w/o $R_{llm}$ & \textbf{1.000} & \textbf{0.384} & 0.509 & 0.102 & 0.744 & \textbf{1.000} & \textbf{0.000} & \textbf{0.955} & 3.859 & \textbf{0.611} \\
\rowcolor{pastelblue}GRPO w/ $R_{llm}$ & \textbf{1.000} & 0.454 & \textbf{0.309} & \textbf{0.110} & \textbf{0.737} & 0.992 & \textbf{0.000} & \textbf{0.955} & \textbf{3.942} & 0.660 \\
\midrule
HiPLEX w/o $R_{llm}$ & \textbf{1.000} & 0.338 & 0.364 & 0.108 & \textbf{0.754} & \textbf{1.000} & \textbf{0.000} & 0.952 & 2.332 & 0.479 \\
\rowcolor{pastelblue}HiPLEX w/ $R_{llm}$ & \textbf{1.000} & \textbf{0.306} & \textbf{0.127} & \textbf{0.134} & 0.734 & \textbf{1.000} & \textbf{0.000} & \textbf{0.960} & \textbf{4.083} & \textbf{0.439} \\
\bottomrule
\end{tabular}%
}
\vspace{-5pt}
\caption{Matched 100-epoch ablation of $R_{llm}$ with the $0$--$2$ rubric. The reported \method{} checkpoint is reused. Best values within each method are bold.}
\label{tab:rllm_abl}
\vspace{-10pt}
\end{table}

\vspace{0pt}
Table~\ref{tab:rllm_abl} compares each training objective with and without semantic feedback.
Adding this reward substantially improves judged interruption-response quality for \method{}, while GRPO shows only a small quality gain. For \method{}, semantic feedback also reduces CANDOR pause and backchannel takeover rates and shortens post-interruption response latency. In this ablation, \method{} improves response quality and interaction timing together when semantic feedback is added.
Appendices~\ref{app:rubric} and \ref{app:semmax} discuss the coarse rubric and semantic overweighting.
These analyses show that rubric choice and semantic-reward strength affect the balance of judged quality and participation, with stronger settings sometimes raising conditional quality as response rates fall.

\subsection{Why Event-Causal Masks Matter}
\vspace{-4pt}
\label{sec:maskabl}
\begin{wraptable}{r}{0.48\linewidth}
\centering
\vspace{-12pt}
\centering
\small
\resizebox{\linewidth}{!}{%
\begin{tabular}{l cccc}
\toprule
Credit rule & Pause$\downarrow$ & Int. lat.$\downarrow$ & Judge$\uparrow$ & Turn$\uparrow$ \\
\midrule
flat GRPO & 0.454 & 0.660 & 3.942 & 0.992 \\
factorized, all frames & 0.264 & 0.874 & 4.330 & 0.958 \\
fixed window & 0.463 & 0.465 & 4.273 & 0.933 \\
random, matched sparsity & 0.028 & 3.277 & 3.579 & 0.067 \\
\rowcolor{pastelblue}event-causal (\method{}) & 0.306 & 0.439 & 4.083 & 1.000 \\
\bottomrule
\end{tabular}%
}
\vspace{-7pt}
\caption{Moshi credit-routing ablation. Factorized rows differ only in credit eligibility. Int. lat. denotes post-interruption response latency. Random masking is excluded
from ranking because its $0.067$ turn rate indicates silence.}
\vspace{-10pt}
\label{tab:maskabl}

\end{wraptable}

Table~\ref{tab:maskabl} demonstrates results where we keeps the factorized policy and rewards fixed while changing timing-credit eligibility. All-frame credit reduces the turn-response rate and increases post-interruption response latency. Matched-random sparsity collapses toward silence, answering only $6.7\%$ under the official turn-taking rule. Thus neither factorization nor sparsity alone explains the result. Event-causal routing is the variant that reaches an official turn TOR of $1.0$ while remaining competitive on pause handling, post-interruption response latency, and quality. Appendix~\ref{app:masks} gives the detailed experiments. The mask definitions illustrate the credit-assignment principle: a delayed-response penalty targets the wait before speech while leaving the eventual onset untouched by that component.

\vspace{-8pt}
\section{Conclusion}
\vspace{-10pt}
We presented \method{}, a hierarchical policy factorization that treats full-duplex speech RL as credit assignment over when to speak and what to say. The policy separates per-frame emission control from conditional content selection, and event-causal masks place timing feedback on decisions tied to generated speech episodes. Compared with our reproduced GRPO baseline, \method{} improves pause and backchannel restraint in both model families and shortens post-interruption response latency on Moshi. The ablations further show that event-causal timing credit shapes turn-taking and interruption behavior, while semantic feedback can improve judged response quality alongside timing. Training sensitivity shows that reward balance and learning rate affect whether the model remains responsive. Event-level success alone can obscure early speech and excessive silence, so evaluation also needs response activity and comparisons with human timing distributions.

\section{Ethics Statement}

Our training data consists of consented dyadic recordings, from which we use only audio and interaction annotations. We do not perform speaker identification, and no raw participant audio is included in the released artifacts.

\section{Reproducibility Statement}

Appendix~\ref{app:hparams} lists the complete training configuration for both reported models, Appendix~\ref{app:data} gives the corpus filtering and event-window extraction rules, and Appendix~\ref{app:algorithm} states the training loop. The evaluation protocol, including the two metric caveats we document, is described in Appendices~\ref{app:pause}, \ref{app:turn}, and \ref{app:statistics}.

\section{AI Use Statement}

We used generative AI tools, including language models and coding agents, to assist with translation, language editing, SVG file generation,
and coding tasks related to research implementation and manuscript preparation, such as debugging,
LaTeX formatting, and consistency checks. The authors reviewed and verified all AI-assisted outputs and take full
responsibility for the experimental design, results, analysis, claims, and final manuscript.
\bibliography{iclr2027_conference}
\bibliographystyle{iclr2027_conference}
\newpage
\appendix

\section{Extended Related Work}
\label{app:related}

\paragraph{Speech language models and full-duplex dialogue.}
Discrete audio tokenization made it possible to model speech with the same machinery as text, first for unconditional
generation \citep{borsos2023audiolm} and then for instruction following and dialogue
\citep{zhang2023speechgpt,nguyen2025spiritlm}. These models generate speech and text in a single token sequence rather than modeling both speakers
as concurrent streams. Full-duplex systems remove that
assumption by generating and listening on the same clock. Moshi does this with an inner-monologue text stream aligned to
streaming audio \citep{defossez2024moshi}, and PersonaPlex builds prompt-based voice and role control on the same
architecture \citep{roy2026personaplexvoicerolecontrol}. Related work targets particular pieces of the same problem:
listening while speaking \citep{ma2024lslm}, backchannels and interruption handling
\citep{veluri2024syncllm}, and broader multimodal interaction \citep{chen2025minmomultimodallargelanguage}.
Our contribution is orthogonal to the architecture: we assume a Moshi-style model and change how RL assigns credit
inside it.

\paragraph{Turn-taking as a modeling problem.}
Long before speech LLMs, dialogue research treated turn-taking as its own prediction task, with models that decide when
a turn is complete and when to take the floor \citep{skantze2021review}. TurnGPT predicts turn completion from text
\citep{ekstedt2020turngpt}, and Voice Activity Projection learns the near-future activity of both speakers directly from
audio \citep{ekstedt2022vap}. That literature makes the point our reward decomposition rests on: turn-taking behaviors
are not a single skill but several, with hesitation, backchanneling, and floor transfer having different cues and
different failure modes. Where those systems predict turn-taking as an external classifier, we optimize it inside the
generative policy itself.

\paragraph{Policy optimization for language models.}
The standard recipe optimizes a language policy against a learned or rule-based reward with PPO-style clipping
\citep{schulman2017ppo,christiano2017preferences,ouyang2022instructgpt}. Two later directions matter here. Preference
optimization removes the reward model entirely \citep{rafailov2023dpo}, which is convenient for pairwise judgments but
ill-suited to our setting, where the signals are measurable event-level quantities such as onset latency rather than
preferences over completions. Group-relative optimization removes the critic instead, standardizing rewards across a
sampled group \citep{shao2024deepseekmath}, which is the estimator we build on: it is stable at our scale and requires no
value network over frame-level speech states. What both share is a flat view of the trajectory, in which one advantage
multiplies every token log-probability. That is exactly the assumption \method{} breaks.

\paragraph{Hierarchy and credit assignment in RL.}
Hierarchical RL introduces temporal abstraction so that decisions at different time scales are learned by different
components: options and their termination conditions \citep{sutton1999between,bacon2017option}, subgoal-conditioned
controllers \citep{kulkarni2016hierarchical,nachum2018hiro}, learned manager-worker splits
\citep{vezhnevets2017feudal}, and value decompositions over a task hierarchy \citep{dietterich2000maxq}. A parallel line
asks not who acts but which past decisions deserve the reward: return decomposition redistributes delayed rewards to the
steps that caused them \citep{arjonamedina2019rudder}, and hindsight credit assignment reweights actions by how much they
actually mattered for the observed outcome \citep{harutyunyan2019hindsight}. HiPER brings this to LLM agents by aligning
advantages with subgoal segments \citep{peng2026hiper}. \method{} sits at the intersection: the hierarchy is not learned
but read exactly out of the pretrained policy's own token partition, and the credit assignment is not learned either but
derived from the audio the policy produced, which makes it an objective-level modification of GRPO rather than a new
architecture.

\paragraph{Evaluating full-duplex behavior.}
Because the failure modes are temporal, they are invisible to text-based dialogue metrics, which has produced benchmarks
that score interaction dynamics directly on streaming audio \citep{lin2025fdbench,lin2025fdbenchv15}. We
evaluate on v1 and, as Appendices~\ref{app:pause} and \ref{app:turn} document, we found two of the v1 metrics to be
degenerate under the official protocol; we report them for comparability but base our claims on the subsets that
discriminate.

\section{Algorithm}
\label{app:algorithm}

Algorithm~1 states the full \method{} training loop. Steps 1--3 are the rollout and grounding stage, steps 4--7 build the
two advantage streams, and steps 8--9 apply them to the two factors of the policy. Everything except the steps marked
\emph{(\method)} is standard group-relative policy optimization~\citep{shao2024deepseekmath}.

\begin{figure}[ht]
\centering
\begin{minipage}{\linewidth}
\rule{\linewidth}{0.8pt}\\[2pt]
\textbf{Algorithm 1}\quad \method{} training loop for one iteration\\[-6pt]
\rule{\linewidth}{0.4pt}
\begin{small}
\textbf{Require:} policy $\pi_\theta$, frozen reference $\pi_{\mathrm{ref}}$, group size $G$, event windows $\mathcal{D}$, clip $\epsilon$, factor weights $\alpha_c,\alpha_l$, KL weights $\beta_c,\beta_l$, learning rate $\eta$
\begin{enumerate}\setlength{\itemsep}{1pt}\setlength{\parskip}{0pt}
\item Sample a window $(s_0,\text{event }e)\sim\mathcal{D}$ with axis $\in\{\text{pause},\text{turn},\text{bc},\text{int}\}$.
\item Stream $G$ causal rollouts from $\pi_{\theta_{\mathrm{old}}}$; at each frame $t$ record the group label $c_t\in\{\pad,\epad,\con\}$ and, when $c_t=\con$, the content token $w_t$.
\item Decode the generated audio, extract speech spans with Silero VAD, and merge them into \emph{speech episodes}, labelling each short (backchannel-like) or sustained (turn-like).
\item For each reward component $k$ of this axis: score every rollout against $e$ and normalize within the group, $A^{(k)}_i=(r^{(k)}_i-\mathrm{mean}_j r^{(k)}_j)/\max(\mathrm{std}_j r^{(k)}_j,\delta)$. If the group is degenerate ($\mathrm{std}_j r^{(k)}_j\approx 0$ with all rollouts failing), re-anchor $A^{(k)}$ against a fixed failure baseline.
\item {(\method)} Build the component-specific event-causal mask $M^{(k)}$ from the generated episodes and the outcome sign: e.g., a successful onset selects its first frame, a late onset selects the pre-onset control interval, an intrusion selects realized speech, and correct silence selects the annotated hesitation or barge-in interval. Appendix~\ref{app:masks} specifies every component and fallback. Stop gradients through episode analysis, reward sign, and mask construction.
\item {(\method)} Route timing credit to control decisions only: $A^{\mathrm{ctrl}}_t=\sum_k w_kA^{(k)}M^{(k)}_t$ (Eq.~\ref{eq:routing}).
\item Score the ASR transcript with the LLM judge, normalize within the group to get $A^{(\mathrm{sem})}$, and {(\method)} route it to the content factor only: $A^{\con}_t=A^{(\mathrm{sem})}\textbf{1}[c_t=\con]$.
\item Form the per-factor ratios $\rho^{\mathrm{ctrl}}_t=\pi^{\mathrm{ctrl}}_\theta(c_t\mid s_t)/\pi^{\mathrm{ctrl}}_{\theta_{\mathrm{old}}}(c_t\mid s_t)$ and $\rho^{\con}_t=\pi^{\con}_\theta(w_t\mid s_t,\con)/\pi^{\con}_{\theta_{\mathrm{old}}}(w_t\mid s_t,\con)$, and the clipped losses $\mathcal{L}_z=-\mathbb{E}[\min(\rho^z_tA^z_t,\mathrm{clip}(\rho^z_t,1\!-\!\epsilon,1\!+\!\epsilon)A^z_t)]$, each averaged over \emph{its own} decisions.
\item Update $\theta\leftarrow\theta-\eta\nabla_\theta[\alpha_c\mathcal{L}_{\mathrm{ctrl}}+\alpha_l\mathcal{L}_{\con}+\beta_c\mathrm{KL}(\pi^{\mathrm{ctrl}}_\theta\|\pi^{\mathrm{ctrl}}_{\mathrm{ref}})+\beta_l\mathrm{KL}(\pi^{\con}_\theta\|\pi^{\con}_{\mathrm{ref}})]$ at the method's tuned rate.
\end{enumerate}
\end{small}
\rule{\linewidth}{0.8pt}
\end{minipage}
\label{alg:hiplex}
\end{figure}

\begin{figure*}[t]
\centering
\includegraphics[width=\textwidth]{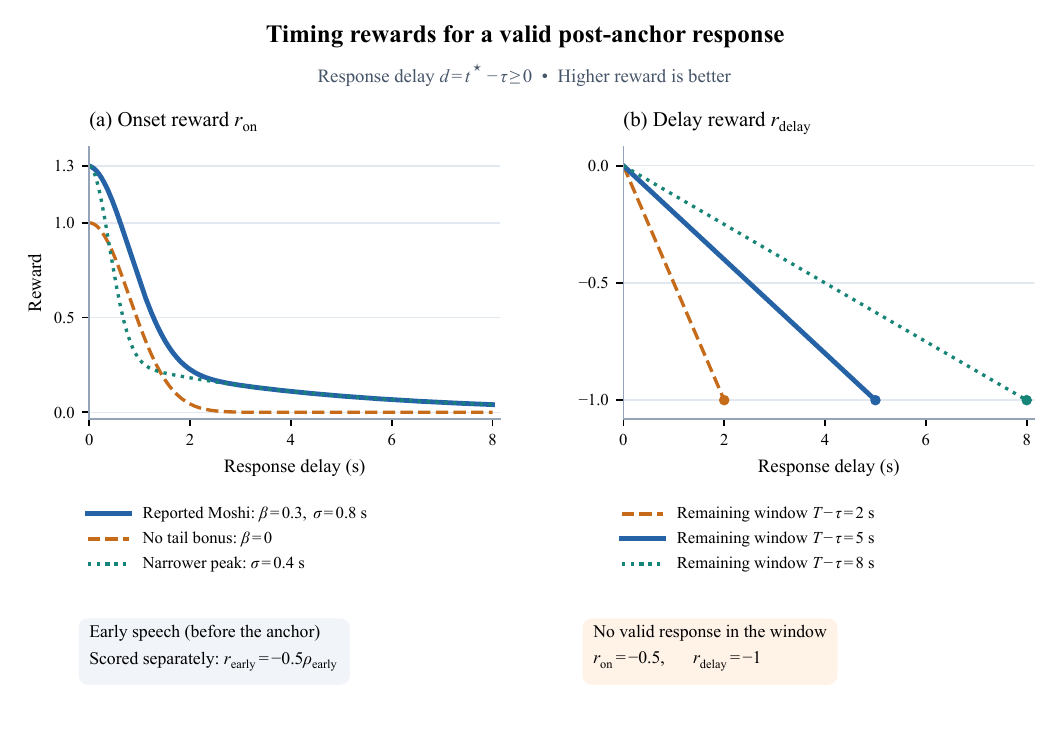}
\caption{Raw timing rewards before group normalization. Both panels show valid responses starting at or after
 the event anchor. \textbf{Left:} the reported Moshi onset reward ($\beta=0.3$, $\sigma=0.8$\,s, $\kappa=4$\,s)
 favors prompt responses and retains a graded tail for late responses. The other curves vary one parameter at a time.
 \textbf{Right:} the delay reward decreases linearly with the fraction of the remaining window spent waiting.
 Each curve ends at its window boundary, where the score reaches $-1$.
 Early speech and missing responses are scored separately as shown below the plots.
 The early-speech penalty shown here applies to turn taking.
 The two components are normalized independently rather than added as raw scores.}
\label{fig:timing-score-shapes}
\end{figure*}
\paragraph{Why two scores for one onset?}
This is a reward-shaping design choice, not a claim that onset and delay are independent behavioral variables. The
onset-shaped component concentrates preference near the anchor and, through its graded tail and fixed miss penalty,
distinguishes a late valid response from a miss; it is also the component re-anchored when every rollout misses. The delay component supplies a dense linear ordering among late
responses and makes delays comparable across window lengths. Crucially, training standardizes the two components
\emph{separately} within each rollout group and only then forms $w_{\mathrm{on}}A^{(\mathrm{on})}+
w_{\mathrm{delay}}A^{(\mathrm{delay})}$. A single raw-sum curve would therefore not depict the optimized signal: its
shape depends on the other rollouts through both means and standard deviations. Figure~\ref{fig:timing-score-shapes}
instead plots the two raw quantities that are actually computed before group normalization.
The distinction between early speech and post-anchor delay is motivated by natural turn exchanges rather than by fitting the reward constants. In CANDOR,
$47.9\%$ of speaker transitions are negative overlaps and $52.1\%$ are positive gaps; the overall median is $+80$\,ms,
and both sides are overwhelmingly shorter than one second \citep{reece2023candor}. This supports treating overlap and
gap as distinct near-zero regimes, but does not imply that all overlap is undesirable. Short acknowledgments are handled
by the backchannel axis; on a turn-taking window, sustained speech before the annotated floor release is instead the
early-start event. Thus CANDOR provides qualitative motivation for this distinction, not a fitted justification for
the penalty magnitude.
Two properties are worth restating here. The control and content log-probabilities in step 8 come from the same softmax
by the exact factorization of \S\ref{sec:factorization}, so no second head is trained and the product of the two factors
is the original token probability. And because the masks in step 5 select approximately $20$--$30\%$ of the frames that a flat
update would touch, the step size is tuned separately to compensate for the lower update frequency, as described in
\S\ref{sec:objective}.

\paragraph{Worked factorization example.}
At one audio frame $t$, suppose the only available tokens are $\pad$, $\epad$, and three content tokens with logits
\begin{align*}
\ell_t(\pad)&=2.0, & \ell_t(\epad)&=0.5,\\
\ell_t(\texttt{hello})&=1.0, & \ell_t(\texttt{yes})&=0.7,
& \ell_t(\texttt{thanks})&=0.2.
\end{align*}
The grouped control logits are therefore
\begin{align*}
&g_t(\pad)=\log\exp(2.0)=2.0, \\
&g_t(\epad)=\log\exp(0.5)=0.5,\\
&g_t(\con)=\log\!\left(\exp(1.0)+\exp(0.7)+\exp(0.2)\right)
          =\log(5.95)\approx1.78.
\end{align*}
Applying the three-way softmax gives
\[
\pi^{\mathrm{ctrl}}_\theta(c_t\mid s_t)
=\frac{\exp g_t(c_t)}{\exp g_t(\pad)+\exp g_t(\epad)+\exp g_t(\con)}
=\begin{cases}
0.49, & c_t=\pad,\\
0.11, & c_t=\epad,\\
0.40, & c_t=\con.
\end{cases}
\]
Thus, $g_t(c)$ is a group score rather than a probability: the $\con$ score pools the support for every content token.
The final $0.40$ is exactly the original softmax probability mass of those three content tokens,
\[
\pi^{\mathrm{ctrl}}_\theta(\con\mid s_t)
=\frac{\exp(1.0)+\exp(0.7)+\exp(0.2)}
{\exp(2.0)+\exp(0.5)+\exp(1.0)+\exp(0.7)+\exp(0.2)}
\approx0.40.
\]
More generally, the outer logarithm in $g_t(c)$ cancels with the exponential in the control softmax:
\[
\pi^{\mathrm{ctrl}}_\theta(c_t=c\mid s_t)
=\frac{\exp g_t(c)}{\sum_{c'}\exp g_t(c')}
=\frac{\sum_{v\in\mathcal{V}_c}\exp\ell_t(v)}
{\sum_{u\in\mathcal{V}}\exp\ell_t(u)}.
\]
Thus, grouping with $g_t(c)=\log\sum_{v\in\mathcal{V}_c}\exp\ell_t(v)$ is exactly a convenient way to express the
original token softmax probability mass assigned to each control group.

\paragraph{Worked onset and delay example.}
Consider a turn opportunity at $\tau=5$\,s in a window ending at $T=10$\,s. For the Moshi reward settings, take
$\sigma=0.8$\,s, $\kappa=4$\,s, and $\beta=0.3$. The outcomes show the localized onset preference, linear delay
penalty, and explicit missing-response values:
\begin{center}
\small
\begin{tabular}{lccc}
\toprule
Outcome & Actual onset $t^\star$ & $r^{(\mathrm{on})}$ & $r^{(\mathrm{delay})}$ \\
\midrule
Starts at the anchor & $5.0$\,s & $1+0.3=1.300$ & $0$ \\
Starts $0.4$\,s late & $5.4$\,s & $e^{-0.125}+0.3e^{-0.1}\approx1.154$ & $-0.4/5=-0.080$ \\
Starts $3.0$\,s late & $8.0$\,s & $e^{-7.03125}+0.3e^{-0.75}\approx0.143$ & $-3/5=-0.600$ \\
No valid response & $\bot$ & $-0.500$ & $-1.000$ \\
\bottomrule
\end{tabular}
\end{center}
The onset-shaped score falls rapidly near the target but retains a positive tail for a late valid response; the delay
score declines linearly and reaches $-1$ at the window horizon. Speech beginning before $\tau$ does not qualify as
$t^\star$ and is handled by the separate early-start component.

\paragraph{Worked event-causal routing example.}
Equation~\ref{eq:routing} uses $k$ to index reward components (such as turn onset or turn delay), not the control
groups $\pad$, $\epad$, and $\con$. For the examples below, we denote frames by their start times in seconds. Suppose the user finishes at $\tau=5.0$\,s and frames arrive every $0.08$\,s.
The following cases illustrate where each normalized advantage is allowed to land.
\begin{itemize}\setlength{\itemsep}{3pt}
\item \textbf{Well-timed turn.} The agent waits until a sustained speech episode starts at $5.28$\,s. If the normalized turn-onset advantage is $A^{(\text{turn onset})}=+1.2$, its mask
selects only the first $80$\,ms onset frame:
\[
M_t^{(\text{turn onset})}=\begin{cases}1,&t=5.28\text{ s},\\0,&\text{otherwise.}\end{cases}
\qquad
A_t^{\mathrm{ctrl}}=\begin{cases}+1.2,&t=5.28\text{ s},\\0,&\text{otherwise.}\end{cases}
\]
The positive signal therefore makes the decision to start speaking at the right time more likely, without rewarding
the earlier waiting frames or every later content token.

\item \textbf{Late turn.} The agent instead remains in the pre-onset control interval from $5.00$ through $7.92$\,s and starts speaking
at $8.00$\,s.  If the turn-delay advantage is $A^{(\text{turn delay})}=-0.8$, the mask selects the preceding waiting
decisions, but \emph{not} the eventual onset. Under the interval-to-frame mapping in Appendix~\ref{app:masks},
the first selected frame starts at $4.96$\,s and contains the anchor at $5.00$\,s:
\begin{align*}
M_t^{(\text{turn delay})}
&=\begin{cases}
1,&4.96\leq t\leq7.92\text{ s},\\
0,&\text{otherwise, including }t=8.00\text{ s},
\end{cases}\\
A_t^{\mathrm{ctrl}}
&=\begin{cases}
-0.8,&4.96\leq t\leq7.92\text{ s},\\
0,&\text{otherwise.}
\end{cases}
\end{align*}
The negative signal says ``do not keep waiting this long,'' rather than ``do not start speaking,'' which would teach
the wrong behavior.

\item \textbf{Overlapping components.} If a control decision at frame $t$ is causal for two components, their routed
advantages add.  For example, with $A^{(1)}=+1.0$, $M_t^{(1)}=1$, $A^{(2)}=-0.3$, and $M_t^{(2)}=1$,
\[
A_t^{\mathrm{ctrl}}=(+1.0)(1)+(-0.3)(1)=+0.7.
\]
If either mask is zero, that component contributes nothing at that frame.
\end{itemize}

\section{Verifying the Exact Policy Factorization}
\label{app:rldiag}

\begin{tcolorbox}[colback=gray!10, colframe=black!70!white]
The control--content factorization preserves the token distribution exactly in real arithmetic. The numerical checks agree up to floating-point roundoff. This identity permits separate objectives for the two factors, but does not make their combined objective equivalent to GRPO.
\end{tcolorbox}

\paragraph{Exact probability reconstruction.}
Fix a state $s$ and use the vocabulary partition defined in Section~\ref{sec:factorization}. For readability, we suppress the conditioning on $s$ below. The singleton groups contain only \pad{} and \epad{}, respectively. Their conditional token probabilities are therefore one. Using the control probabilities and the content conditional in Equation~\eqref{eq:factorization}, the reconstructed token log-probability is
\[
L_\theta(v)=
\begin{cases}
\log\pi_\theta^{\mathrm{ctrl}}(\pad), & v=\pad,\\
\log\pi_\theta^{\mathrm{ctrl}}(\epad), & v=\epad,\\
\log\pi_\theta^{\mathrm{ctrl}}(\con)
+\log\pi_\theta^\con(v\mid\con), & v\in\mathcal V_\con.
\end{cases}
\]
Substituting the grouped softmax definitions cancels the within-group normalizer and gives
\[
L_\theta(v)=\log\pi_\theta(v).
\]
This identity requires both factors to be derived from the same token distribution. To reconstruct sampling probabilities, any temperature scaling or support restriction must also be included in that distribution.

\paragraph{Numerical check.}
The diagnostic computes the absolute reconstruction error for each tested token:
\[
\epsilon(v)=\bigl|\log\pi_\theta(v)-L_\theta(v)\bigr|.
\]
In the saved diagnostic report, the median error is zero for all three input dtypes. The maximum errors are
\[
\begin{array}{c|ccc}
\text{Input dtype} & \texttt{float32} & \texttt{bfloat16} & \texttt{float16}\\
\hline
\max\epsilon & 7.6\times10^{-6} & 3.8\times10^{-6} & 3.8\times10^{-6}
\end{array}
\]
The diagnostic casts each input to its stated dtype, then computes both log-probabilities in \texttt{float32}.

\paragraph{Exact KL decomposition.}
For two policies with the same vocabulary partition and positive token probabilities, the KL chain rule gives
\[
\begin{aligned}
\mathrm{KL}(\pi_\theta\|\pi_{\mathrm{ref}})
&=\mathrm{KL}(\pi_\theta^{\mathrm{ctrl}}\|\pi_{\mathrm{ref}}^{\mathrm{ctrl}})\\
&\quad+\pi_\theta^{\mathrm{ctrl}}(\con)\,
\mathrm{KL}\!\left(\pi_\theta^\con(\cdot\mid\con)
\middle\|\pi_{\mathrm{ref}}^\con(\cdot\mid\con)\right).
\end{aligned}
\]
The singleton groups contribute zero conditional KL. The content KL is weighted by the current policy's probability of selecting content. The saved diagnostic compares this decomposition against token KL using a perturbed copy of the input logits, with a maximum absolute discrepancy of $5.8\times10^{-7}$. A separately weighted sum of control and content KL penalties, as used in our objective, is not generally equal to the full token KL.

\paragraph{Relation to GRPO.}
GRPO~\citep{shao2024deepseekmath} uses a shared advantage for the sampled token decisions in a rollout. Differentiating the reconstruction identity gives
\[
\nabla_\theta\log\pi_\theta(v)=\nabla_\theta L_\theta(v).
\]
Thus, for a content token, the control and conditional-content score gradients sum to the full token score gradient. For \pad{} and \epad{}, only the control term is needed. The same on-policy score-function estimator is recovered when these terms receive the same fixed advantage, equal weights, identical token eligibility, and a common normalization over sampled tokens.

This gradient identity does not imply equality of the separately clipped objectives. In particular, for a content token the full importance ratio is a product:
\[
\frac{\pi_\theta(v)}{\pi_{\theta_{\mathrm{old}}}(v)}
=
\frac{\pi_\theta^{\mathrm{ctrl}}(\con)}{\pi_{\theta_{\mathrm{old}}}^{\mathrm{ctrl}}(\con)}
\frac{\pi_\theta^\con(v\mid\con)}{\pi_{\theta_{\mathrm{old}}}^\con(v\mid\con)}.
\]
Even without clipping, separate factor ratios need not give the same gradient as this product away from the sampling policy. Our factor-specific advantages, masks, normalization, clipping, and KL penalties further distinguish the objective from GRPO.

\section{Reward Components: Definitions, Weights, and Edge Cases}
\label{app:rewards}

\paragraph{Component index and normalization.}
\begin{itemize}
\item \textbf{Component index.} In Eq.~\ref{eq:group-advantage}, $k$ identifies one scalar timing score
for the interaction axis of the sampled window. For example, a turn-taking rollout receives separate onset,
delay, and early-start scores. The rollout index $i$ and control-action label are separate indices.
\item \textbf{Within-window normalization.} Each component is standardized across the $G$ rollouts
of the same window. The implementation uses the population standard deviation,
$s_r^{(k)}=\sqrt{G^{-1}\sum_{i=1}^{G}(r_i^{(k)}-\bar r^{(k)})^2}$,
and divides by $\max(s_r^{(k)},\delta)$, with $\delta$ as specified in the main text.
A constant component therefore has zero advantage before the sparse-failure exception below.
\item \textbf{Active components.} Table~\ref{tab:reward-components} lists the active timing-score
inventory and the weights used by the reported Moshi and PersonaPlex runs. Only components with nonzero
weight contribute to the control update.
\item \textbf{Semantic score.} The semantic score is normalized separately and routed to the conditional
content policy; it is not part of this timing-component index.
\end{itemize}

\begin{table}[ht]
\centering
\small
\setlength{\tabcolsep}{4pt}
\begin{tabular}{lllcc}
\toprule
Window axis & Reward & Implementation name & Moshi & PersonaPlex \\
\midrule
Turn taking & $r^{(\mathrm{on})}$ & \texttt{turn\_onset} & 1 & 1 \\
 & $r^{(\mathrm{delay})}$ & \texttt{turn\_delay} & 1 & 2 \\
 & $r^{(\mathrm{early})}$ & \texttt{turn\_early\_start} & 1 & 1 \\
\midrule
Pause & $r^{(\mathrm{intr})}$ & \texttt{interaction} & 1 & 1 \\
 & $r^{(\mathrm{frac})}$ & \texttt{listen\_false\_alarm} & 1 & 1 \\
\midrule
Backchanneling & $r^{(\mathrm{F1})}$ & \texttt{interaction} & 1 & 1 \\
 & $r^{(\mathrm{time})}$ & \texttt{bc\_timing} & 1 & 1 \\
 & $r^{(\mathrm{miss})}$ & \texttt{bc\_miss} & 1 & 1 \\
 & $r^{(\mathrm{false})}$ & \texttt{bc\_false\_alarm} & 0.5 & 0.5 \\
 & $r^{(\mathrm{long})}$ & \texttt{bc\_overlong} & 1 & 1 \\
\midrule
Interruption & $r^{(\mathrm{yield})}$ & \texttt{interruption\_yield} & 2 & 0.5 \\
 & $r^{(\text{re-onset})}$ & \texttt{reentry\_onset} & 1 & 1 \\
 & $r^{(\text{re-delay})}$ & \texttt{reentry\_delay} & 1 & 2 \\
\bottomrule
\end{tabular}
\caption{Active timing components and their weights $w_k$ in the reported training configurations.
Turn taking, pause, backchanneling, and interruption use 3, 2, 5, and 3 components, respectively.
The component labels $\text{re-onset}$ and $\text{re-delay}$ denote re-entry onset and delay after an interruption.
The implementation name \texttt{interaction} refers to different scores on different axes.}
\label{tab:reward-components}
\end{table}

The weights multiply the independently standardized advantages, rather than the raw scores before normalization:
$A_{i,t}^{\mathrm{ctrl}}=\sum_{k\in\mathcal{K}(e)}w_k A_i^{(k)}M_{i,t}^{(k)}$,
where $\mathcal{K}(e)$ contains the active components for the window's axis $e$.
For example, Moshi turn-taking uses the three separately normalized advantages with unit weights, whereas
PersonaPlex doubles the delay advantage. Raw penalties such as $-0.5$ below define the scores $r_i^{(k)}$;
they are distinct from the weights $w_k$ in the table. All timing scores are oriented so that larger is better.

\paragraph{Speech spans and event anchors.}
In the training implementation, Silero VAD~\citep{silero2024vad} extracts speech spans $U$ from the generated audio.
Let $S$ be the speech episodes obtained by merging these spans separated
by at most $1$\,s. Episodes of duration at most $1$\,s are short; longer episodes are sustained.
Episode duration and onset determine event detection, while speech-fraction and overlap penalties sum the durations
of the underlying spans $U$, without filling the gaps introduced by episode merging.
Let $T$ be the window end and define
$D_U(a,b)=\sum_{u\in U}\max(0,\min(u_{\mathrm{end}},b)-\max(u_{\mathrm{start}},a))$.
All times are relative to the window start. Duration denominators are lower-bounded by $10^{-6}$\,s in the implementation.

\paragraph{Turn taking.}
For turn taking, $\tau$ is the user's turn end and $t^\star$ is the first sustained episode onset at or after $\tau$.
\begin{itemize}
\item \textbf{Onset ($r^{(\mathrm{on})}$).} The onset score in Eq.~\ref{eq:onset}, with $\sigma=0.8$\,s,
$\kappa=4$\,s, and onset bonus $\beta=0.3$ for Moshi or $0.5$ for PersonaPlex.
A missing qualifying response receives $-0.5$. The exponential bonus retains a response signal beyond the
narrow Gaussian peak.
\item \textbf{Delay ($r^{(\mathrm{delay})}$).} The delay score in Eq.~\ref{eq:onset}, the negative delay divided by
the remaining window duration and clipped to $[-1,0]$. A missing response receives $-1$ for a positive
remaining horizon.
\item \textbf{Early start ($r^{(\mathrm{early})}$).} The pre-anchor speech penalty is
\[
 r^{(\mathrm{early})}
 =-0.5\min\!\left(1,\frac{D_U(0,\tau)}{\max(\tau,10^{-6})}\right).
\]
\end{itemize}
The raw negative delay in seconds is also logged as \texttt{interaction}, with $-(T-\tau)$ for a missing
response. Its weight is zero, so it does not add a fourth turn-taking advantage.

\paragraph{Pause.}
\begin{itemize}
\item \textbf{Intrusion ($r^{(\mathrm{intr})}$).} A binary penalty detects sustained speech:
\[
 r^{(\mathrm{intr})}=-\mathbf{1}[\exists s\in S:\ |s|\geq 1\,\mathrm{s}].
\]
\item \textbf{Speech fraction ($r^{(\mathrm{frac})}$).} A duration-based penalty also discourages shorter intrusions:
\[
 r^{(\mathrm{frac})}
 =-0.5\min\!\left(1,\frac{\sum_{u\in U}|u|}{\max(T,10^{-6})}\right).
\]
\end{itemize}
Both scores cover the sampled pause-handling window. Their frame-level credit is selected by the masks in
Appendix~\ref{app:masks}.

\paragraph{Backchanneling.}
Let $W=\{W_j\}_{j=1}^{m}$ be the annotated opportunity windows, $B$ the short episodes, and $L$ the sustained
episodes. For a short episode $b$, let $d(b,W_j)$ be the distance of its centre to $W_j$, with distance zero
inside the window. The five active components are:
\begin{itemize}
\item \textbf{F1 ($r^{(\mathrm{F1})}$).} The legacy backchannel score is
\[
 r^{(\mathrm{F1})}=\frac{2\mathrm{TP}}{2\mathrm{TP}+\mathrm{FP}+\mathrm{FN}}.
\]
Short episodes are visited in temporal order and matched to the first still-unmatched opportunity within
$1$\,s. Unmatched short episodes and all sustained episodes count as false positives. Unmatched opportunities
count as false negatives. The score is zero when its denominator is zero.
\item \textbf{Timing ($r^{(\mathrm{time})}$).} The soft proximity score is
\[
 r^{(\mathrm{time})}=\frac{1}{m}\sum_{j=1}^{m}\exp\!\left[-2\min_{b\in B}d(b,W_j)\right].
\]
The score is zero if $W$ or $B$ is empty. This is a soft proximity score over all opportunities, without the
hard matching cutoff or a one-to-one constraint.
\item \textbf{Miss ($r^{(\mathrm{miss})}$).} The unmatched-opportunity penalty is
\[
 r^{(\mathrm{miss})}=-0.5\frac{m-n_W}{m},
\]
with score zero when $m=0$. Here $n_W$ is the number of matched opportunities.
\item \textbf{False alarm ($r^{(\mathrm{false})}$).} The unmatched-episode penalty is
\[
 r^{(\mathrm{false})}=-0.5\frac{|B|-n_B}{\max(1,|B|)}.
\]
For these two components, $n_W$ and $n_B$ count matched opportunities and short episodes under a one-to-one
matching: candidate pairs within $1$\,s are sorted by distance, then greedily accepted if neither member is
already matched. This matching is computed separately from the legacy F1 matching above.
\item \textbf{Overlong ($r^{(\mathrm{long})}$).} Sustained episodes incur
\[
 r^{(\mathrm{long})}=-0.5\,\mathbf{1}[|L|>0].
\]
The penalty is binary: one or more sustained episodes incurs the same score, rather than a penalty proportional
to their count or excess duration.
\end{itemize}
Thus F1, proximity, omission, false-alarm, and duration scores each have their own group-relative advantage.

\paragraph{Interruption.}
For a barge-in spanning $[\beta_0,\beta_1]$, re-entry is anchored at $\tau=\beta_1$.
\begin{itemize}
\item \textbf{Yield ($r^{(\mathrm{yield})}$).} The overlap penalty is
\[
 r^{(\mathrm{yield})}
 =-\min\!\left(1,\frac{D_U(\beta_0+0.16,\beta_1)}{1\,\mathrm{s}}\right).
\]
The $0.16$\,s grace is two audio frames. The score is zero if no interruption interval is supplied.
\item \textbf{Re-entry onset ($r^{(\text{re-onset})}$).} This uses the same onset score as turn taking, evaluated at the first sustained
episode starting at or after $\beta_1$. Speech already crossing the anchor is not a new response.
\item \textbf{Re-entry delay ($r^{(\text{re-delay})}$).} This uses the same normalized-delay score as turn taking, now relative to $\beta_1$.
\end{itemize}
The raw negative re-entry delay in seconds is logged as \texttt{interaction} with zero weight.

\paragraph{Degenerate groups.}
\begin{itemize}
\item \textbf{Activation.} The sparse-failure escape activates only if all component scores are finite,
their maximum absolute deviation from the group mean is at most $10^{-8}$, and their maximum is at most
$-0.05$. This uniformity tolerance is a separate check from the denominator floor $\delta$.
\item \textbf{Replacement.} Before weighting and masking, the component advantage is replaced by
$-1.5$ for $r^{(\mathrm{on})}$ and $r^{(\mathrm{miss})}$, or by $-1.0$ for $r^{(\text{re-onset})}$.
This replacement does not change the raw reward.
\item \textbf{Other groups and components.} Ordinary group-relative normalization applies when the
activation conditions fail and to all unlisted components, including zero advantage for constant groups.
\end{itemize}

\section{Event-Causal Masks in Detail}
\label{app:masks}

This appendix specifies the masks that \S\ref{sec:eventcausal} summarizes. Throughout, \emph{speech episodes} are the
Silero VAD speech spans of the agent's own output, merged across gaps of at most $1$\,s and labelled short (backchannel-like) or
sustained (turn-like) by duration. Masks combine these generated episodes with the ground-truth event timing and component advantage, so credit
depends on what the policy actually did in relation to the target event. Reward symbols follow Appendix~\ref{app:rewards}.

\paragraph{Operational meaning of event-causal.}
The term \emph{causal} denotes structural eligibility for policy-gradient credit, not a learned causal graph or a
counterfactual effect estimator. For each reward component we begin with the action type that can directly realize the
event under the model's autoregressive emission process (starting, waiting, continuing, or stopping speech), then select
the realized frames of that type from the generated episode. Proximity alone is insufficient: a frame can lie inside an
annotated event window yet be unable to correct the observed failure. This definition is falsifiable at the mechanism
level---replacing it with the whole window or a random set of equal density gives the controlled comparisons in
Table~\ref{tab:maskabl}.

All episode extraction, reward signs, and binary masks are computed after rollout under
$\pi_{\theta_{\mathrm{old}}}$ and treated as fixed, non-differentiable credit weights during the PPO update. Thus no
gradient is taken through VAD extraction, episode boundaries, reward sign, or mask construction; gradients enter only
through the selected control log-probabilities. If masks from multiple independently normalized reward components select
the same frame, their signed advantages add as in Eq.~\eqref{eq:routing}, after which the control loss is normalized over
the selected decisions.

\paragraph{Episode construction.}
Silero VAD speech spans separated by at most $1$\,s are merged. An episode of duration at most $1$\,s is treated as
short (backchannel-like), and a longer episode as sustained (turn-like). These constants are shared with reward
extraction. The masks below are temporal eligibility regions over sampled control decisions; a waiting interval is not
defined by blindly equating every raw \pad{} token with acoustic silence.

\paragraph{Turn taking.}
Let $\tau$ be the ground-truth end of the user's turn and $t^\star$ the first sustained episode onset at or after
$\tau$. The onset and delay components, $r^{(\mathrm{on})}$ and $r^{(\mathrm{delay})}$, are masked separately using
the sign of their respective advantages.
\begin{itemize}
\item \textbf{Positive onset or delay advantage.} Select the onset interval
$[t^\star,\min(T,t^\star+0.08\,\mathrm{s}))$. Under ordinary group normalization, a positive advantage
indicates a score above the group mean. Both scores evaluate when the response starts, so this local interval
assigns credit to beginning speech without reinforcing the preceding wait.
\item \textbf{Non-positive onset or delay advantage.} Select $[\tau,t^\star)$, excluding the response onset.
A negative advantage then discourages the waiting decisions that delayed the response. Penalizing the eventual
onset could instead suppress the response. A zero advantage makes no contribution to the update.
If the selected interval contains no frames, use the onset interval as a fallback.
\item \textbf{Missing response.} With no valid post-anchor onset, select $[\tau,T)$.
If a sustained episode already crosses $\tau$, select the boundary interval
$[\tau,\min(T,\tau+0.08\,\mathrm{s}))$ and any waiting interval after that episode ends.
When several episodes qualify as crossing the anchor, use the latest end time.
The continuing episode is not counted as a new response.
\item \textbf{Early start.} For $r^{(\mathrm{early})}$, select the portions of generated episodes before $\tau$.
If no episode starts before the anchor, select $[0,\min(\tau,T))$ instead.
\end{itemize}

\paragraph{Pause handling.}
The binary intrusion and speech-fraction components, $r^{(\mathrm{intr})}$ and $r^{(\mathrm{frac})}$, use the same mask.
\begin{itemize}
\item \textbf{Speech present.} Select all generated speech episodes in the sampled pause-handling window.
Their component advantages update these speech decisions without assigning intrusion penalties to unrelated
silent frames.
\item \textbf{Silent rollout.} Select the ground-truth pause windows. A positive advantage then reinforces
remaining silent in those windows. If no pause window is supplied, select the whole segment.
\end{itemize}

\paragraph{Backchanneling.}
Expand each ground-truth opportunity window by $1$\,s on both sides and clip it to $[0,T]$.
For mask construction, a short episode matches when its centre lies in any expanded window.
This membership rule does not require the one-to-one matching used to compute some reward scores.
\begin{itemize}
\item \textbf{Timing and miss ($r^{(\mathrm{time})},r^{(\mathrm{miss})}$).} For each component, a positive
advantage with at least one matching short episode selects those episodes. Otherwise, select the expanded
opportunity windows. A negative miss advantage therefore discourages the decisions made while the opportunity passed.
\item \textbf{False alarm ($r^{(\mathrm{false})}$).} Select generated episodes whose centres lie outside all
expanded opportunity windows. The implementation checks both short and sustained episodes when constructing this mask.
\item \textbf{Overlong ($r^{(\mathrm{long})}$).} If sustained episodes exist, select only the portion of each
after its first second. This targets continuing beyond the allowed duration without penalizing the beginning.
Otherwise, select the local stop interval from $0.08$\,s before to $0.08$\,s after each short episode's end,
clipped to the episode start and segment end. Complete silence has no stop action and receives no duration credit.
\item \textbf{F1 ($r^{(\mathrm{F1})}$).} Use the union of the expanded opportunity windows, unmatched-episode
regions, and duration regions above. This union is constructed without a sign-dependent switch to matching
episodes. The F1 advantage weights the resulting union mask.
\end{itemize}

\paragraph{Interruption handling.}
Let $[\beta_0,\beta_1]$ be the ground-truth interruption interval.
\begin{itemize}
\item \textbf{Yield ($r^{(\mathrm{yield})}$).} If generated episodes overlap the interruption, select their
intersections with $[\beta_0,\beta_1]$. Otherwise, select the entire interruption interval, allowing positive
advantages to reinforce silence. The reward calculation allows a $0.16$\,s grace period, but the mask covers
the overlap with the full interruption interval.
\item \textbf{Re-entry ($r^{(\text{re-onset})},r^{(\text{re-delay})}$).} Apply the turn-taking
onset-versus-waiting rules with anchor $\tau=\beta_1$. The same missing-response and crossing-episode cases apply.
\end{itemize}

\paragraph{Quantization and empty masks.}
A selected interval $[a,b)$ is mapped to frame indices from $\lfloor a/0.08\rfloor$ to
$\lceil b/0.08\rceil-1$, clipped to the rollout length. Thus an $80$\,ms onset interval can cover two frame
indices when the VAD boundary falls between frame boundaries. Very short or clipped regions can select no frame. In that case the implementation selects the component's ordinary
causal anchor so that the reward does not silently disappear. The only exceptions are false-alarm and overlong-duration
components, for which an empty set means that the corresponding action did not occur. If several independently
normalized components select the same frame, their signed advantages add before the factor-specific loss is normalized
by its number of eligible decisions.

\paragraph{Degenerate groups.}
Sparse events produce groups in which every rollout fails identically, most often backchannels early in training. For an
explicitly named component, the escape activates only when all rewards are finite, their maximum deviation from the group
mean is at most $10^{-8}$, and every reward is at most $-0.05$. The otherwise-zero normalized advantage is then replaced
by the bounded constant $-s_k$ at that component's existing event-causal mask. We use
$s_{\mathrm{bc\_miss}}=1.5$, $s_{\mathrm{turn\_onset}}=1.5$, and
$s_{\mathrm{reentry\_onset}}=1.0$; all other components retain ordinary group-relative normalization.

\paragraph{Full controlled routing ablation.}
Table~\ref{tab:maskabl-full} expands the compact main-paper ablation. All factorized rows inherit the same reported
Moshi configuration and 100-epoch endpoint and differ in the temporal eligibility of control credit. The all-frame row
therefore isolates the hierarchy and level-specific routing from event masking, while the matched-random row controls
for sparsity without event alignment.

\begin{table}[h]
\centering
\small
\resizebox{\linewidth}{!}{%
\begin{tabular}{lrrrrrr}
\toprule
Credit rule & Pause$\downarrow$ & BC TOR$\downarrow$ & BC freq$\uparrow$ & Turn$\uparrow$ & Judge$\uparrow$ & Int. lat.$\downarrow$ \\
\midrule
flat token GRPO & 0.454 & 0.309 & 0.110 & 0.992 & 3.942 & 0.660 \\
factorized, all frames & 0.264 & 0.455 & 0.081 & 0.958 & 4.330 & 0.874 \\
factorized, fixed window & 0.463 & 0.382 & 0.117 & 0.933 & 4.273 & 0.465 \\
factorized, matched random & 0.028 & 0.091 & 0.009 & 0.067 & 3.579 & 3.277 \\
\rowcolor{pastelblue}factorized, event-causal & 0.306 & 0.127 & 0.134 & 1.000 & 4.083 & 0.439 \\
\bottomrule
\end{tabular}%
}
\caption{Full Moshi credit-routing ablation. Int. lat. denotes post-interruption response latency. The random row's low pause TOR coincides with near-zero backchannel
frequency and turn response, distinguishing silence from balanced restraint.}
\label{tab:maskabl-full}
\end{table}

\section{Training Data: Subset Selection and Preprocessing}
\label{app:data}

\paragraph{Source corpus and subset selection.} We use the Seamless Interaction corpus~\citep{seamless2025} as the source of two-channel human conversations. The corpus spans a wide range of prompted dyadic interactions, including physical-action and gesture-oriented prompts (e.g., charades-style games) recorded for its video modality; these are irrelevant to speech-only interaction modeling. We therefore keep only the \emph{ipc\_conversation} interaction type from the \emph{naturalistic} and \emph{improvised} subsets, which are free-form and role-play conversation respectively, and discard all action-oriented interaction types. We download only the two per-speaker audio channels with their Silero-VAD interval and transcript annotations; the much larger video and feature archives are never fetched.

\paragraph{Corpus statistics.} After filtering, the training pool contains approximately 22.1k two-channel interactions ($\sim$44.3k channel recordings), with a mean duration of 4.4 minutes per recording, roughly 1.6k hours of dyadic conversation (3.2k channel-hours). Conversations are split into train and validation at the conversation level before any window extraction, so no speaker pair or local context appears on both sides.
Table~\ref{tab:data_stats} summarizes the resulting train and validation splits.
\begin{table}[h]
\centering
\caption{Training-data statistics after conversation-only filtering (\S\ref{app:data}) Window hours count extracted RL windows; both channel orientations of each interaction are used.}
\label{tab:data_stats}
\begin{tabular}{lccc}
\toprule
Split & Channel recordings & Event windows & Window hours \\
\midrule
Train & 44{,}292 & 73{,}561 (18.3--18.4k/axis) & 271 \\
Validation & 1{,}665 & 2{,}048 (512/axis) & 7.5 \\
\bottomrule
\end{tabular}
\end{table}

\paragraph{Event-window extraction.} Using the released VAD intervals, inter-pausal units separated by $\leq$1\,s are merged into utterances, and the four axis-specific RL window types are selected by rule:
\begin{itemize}
    \item \textbf{Pause}: an utterance of $\geq$4\,s containing an internal silent pause; the annotated pause spans become the listen windows.
    \item \textbf{Turn}: consecutive utterances $u,v$ by different speakers, each $\geq$5\,s, with onset gap $\leq$0.4\,s and $v$ extending $>$1\,s beyond $u$; the boundary is the turn-end event.
    \item \textbf{Backchannel}: an utterance of $\geq$5\,s containing embedded other-speaker speech of $\leq$1\,s; the embedded spans become backchannel-opportunity windows.
    \item \textbf{Interruption}: an $x,y,x,y$ speaker pattern of four utterances, each $\geq$3\,s, where the third overlaps the second (barge-in) and the fourth continues $>$1\,s past it (observable re-entry).
\end{itemize}
Both channel orientations of every interaction are emitted (each speaker in turn plays the agent role). The resulting manifest contains 73{,}561 training windows balanced across axes (18.3--18.4k per axis; 271 hours of window audio) and 2{,}048 validation windows (512 per axis; 7.5 hours). The policy input is strictly causal; future annotations are used only for window selection and reward computation (\S4.1).

\section{Full Training Configuration}
\label{app:hparams}

Table~\ref{tab:hparams} lists every setting used for the two \method{} models we report, both trained from the
pretrained checkpoint on Seamless Interaction event windows (Appendix~\ref{app:data}). The two runs share the entire
recipe except for the entries in the lower block, which are the adjustments \S\ref{sec:dynamics} motivates: a halved
step size and a reward balance tilted toward initiation, both needed because PersonaPlex's prompt-conditioned prior
makes the suppression gradient much cheaper to follow.

We use AdamW~\citep{loshchilov2019adamw} with a cosine learning-rate schedule~\citep{loshchilov2017sgdr}.

\begin{table}[h]
\centering
\small
\resizebox{\linewidth}{!}{%
\begin{tabular}{lll}
\toprule
 & Moshi + \method{} & PersonaPlex + \method{} \\
\midrule
\multicolumn{3}{l}{\emph{Model and data}} \\
Pretrained checkpoint & \texttt{kyutai/moshika-pytorch-bf16} & \texttt{nvidia/personaplex-7b-v1} \\
Conditioning & none & text prompt + voice prompt \\
Training windows & \multicolumn{2}{l}{73{,}561 (18.3--18.4k per axis), Seamless Interaction} \\
Frame rate & \multicolumn{2}{l}{$12.5$\,Hz ($0.08$\,s per frame)} \\
Context prepended & \multicolumn{2}{l}{$0$--$30$\,s, probability $0.5$, linear schedule} \\
Training seed & \multicolumn{2}{l}{$42$} \\
\midrule
\multicolumn{3}{l}{\emph{Rollout}} \\
Group size $G$ & \multicolumn{2}{l}{16} \\
Groups per epoch & \multicolumn{2}{l}{32} \\
Text / audio temperature & \multicolumn{2}{l}{$0.7$ / $0.8$, audio top-$k$ $250$} \\
\midrule
\multicolumn{3}{l}{\emph{Objective}} \\
Clip range $\epsilon$ (both factors) & \multicolumn{2}{l}{$0.2$} \\
Factor weights $\alpha_c,\alpha_l$ & \multicolumn{2}{l}{$1.0$, $1.0$} \\
Training ASR / judge & \multicolumn{2}{l}{\texttt{parakeet-tdt-0.6b-v2} / \texttt{gemini-2.5-flash-lite}} \\
Semantic routing & \multicolumn{2}{l}{conditional content factor only} \\
Sparse-failure escape & \multicolumn{2}{l}{bc miss $1.5$, turn onset $1.5$, re-entry onset $1.0$} \\
Optimizer & \multicolumn{2}{l}{AdamW ($\beta{=}0.9/0.95$, weight decay $0.1$), cosine schedule, grad clip $2.0$} \\
Epochs & \multicolumn{2}{l}{100} \\
Hardware & \multicolumn{2}{l}{16$\times$H100, bf16, $\sim$13\,h per 100 epochs} \\
\midrule
\multicolumn{3}{l}{\emph{Family-specific settings}} \\
Learning rate $\eta$ & $4\times10^{-6}$ & $2\times10^{-6}$ \\
Control KL $\beta_c$ & $0.01$ & $0.03$ \\
Content KL $\beta_l$ & $0.02$ & $0.02$ \\
Training-judge rubric & \multicolumn{2}{l}{$0$--$2$} \\
Semantic reward weight & \multicolumn{2}{l}{$1.0$} \\
Axis mix per epoch (pause/turn/bc/int) & uniform & $5/13/6/8$ \\
Turn and re-entry delay weight & $1.0$ & $2.0$ \\
Interruption-yield weight & $2.0$ & $0.5$ \\
Onset response bonus & $0.3$ & $0.5$ \\
\bottomrule
\end{tabular}%
}
\caption{Complete training configuration for the two reported \method{} models. Everything above the last block is
shared; the last block contains the only differences, all of them consequences of the stability analysis in
\S\ref{sec:dynamics}. Remaining reward-shaping constants (post-anchor onset width $0.8$\,s, graded-tail scale $4$\,s,
missing-event and false-alarm penalties $-0.5$, backchannel maximum duration $1.0$\,s, interruption grace $0.16$\,s)
are shared by both runs.}
\label{tab:hparams}
\end{table}

\section{Semantic-Reward and Evaluation-Judge Protocol}
\label{app:judgeprotocol}

The semantic signal used for training and the response-quality evaluator are separate calls with different rubrics.
This section records the exact models and prompts used by the reported runs.

\paragraph{Training reward.}
Both the user context and generated agent audio are transcribed with
\texttt{nvidia/parakeet-tdt-0.6b-v2}~\citep{nvidia2025parakeet}. The reported Moshi run queries
\texttt{gemini-2.5-flash-lite}~\citep{google2025geminiflashlite} once per response with temperature $0$ and at most 64 output tokens. Both reported Moshi
and PersonaPlex runs use the same $0$--$2$ rubric. The literal suffix \texttt{/no\_think} is appended to the following
system prompt to disable thinking:

\begin{figure}[h!]
    \centering
    \begin{minipage}{\textwidth}
    \fbox{
        \begin{minipage}{\textwidth}
            \fontsize{10}{10}\selectfont
            %%%%%%%%%
                % Title
            \textbf{Evaluating dialogue response quality} \\
            \rule{\textwidth}{0.4pt}
            \fontsize{9}{10}\selectfont
            %%%%%%%%%%
                % content start
            You are evaluating a spoken dialogue model's response in a turn-taking scenario.\\
            You will be given a conversation history consisting of alternating USER and MODEL turns.\\
            The dialogue model should respond appropriately given the full conversation history.\\
            Rate the MODEL response on a scale from 0 to 2 using the following criteria:\\[2pt]

            0: Irrelevant, unnatural, or no meaningful response (irrelevant, silence, or gibberish).\\
            1: Loosely or partially related, but generic and lacking specificity to the context.\\
            2: Natural, specific, and coherently grounded in the context.\\[2pt]

            Output only a single integer from 0 to 2.\\
            /no\_think \\
            %%%%%%%%%%%
                % content end
            \rule{\textwidth}{0.4pt}
            \fontsize{9}{10}\selectfont
            %%%%%%%%
                % output
            \textbf{Response}: 0, 1, or 2
        \end{minipage}
        }
    \end{minipage}

    \caption{A prompt for evaluating the quality of a dialogue model response in a conversational turn-taking scenario.}
    \label{app-fig:dialogue-response-evaluation-prompt}
\end{figure}

The user message is formatted as \texttt{USER: <transcript>} followed by
\texttt{MODEL: <transcript>}. Empty, irrelevant, or unintelligible responses are assigned score $0$ by the same rubric
rather than filtered from the rollout group. The integer is standardized only against the other rollouts for the same
context before becoming $A^{(\mathrm{sem})}$.

The separate $0$--$5$ Moshi rubric ablation in Appendix~\ref{app:rubric} uses the following finer prompt; it is not the
prompt used by either reported main model:

\begin{figure}[h!]
    \centering
    \begin{minipage}{\textwidth}
    \fbox{
        \begin{minipage}{\textwidth}
            \fontsize{10}{10}\selectfont
            %%%%%%%%%
                % Title
            \textbf{Evaluating dialogue response quality} \\
            \rule{\textwidth}{0.4pt}
            \fontsize{9}{10}\selectfont
            %%%%%%%%%%
                % content start
            You are evaluating a spoken dialogue model's response in a turn-taking scenario.\\
            You will be given a conversation history consisting of alternating USER and MODEL turns.\\
            The dialogue model should respond appropriately given the full conversation history.\\
            Rate the MODEL response on a scale from 0 to 5 using the following criteria:\\[2pt]

            0: No meaningful response (silence or gibberish).\\
            1: Irrelevant or unnatural response.\\
            2: Barely related, mostly generic filler.\\
            3: Loosely or partially related, but generic and lacking specificity.\\
            4: Relevant and natural, grounded in the context with minor flaws.\\
            5: Natural, specific, and coherently grounded in the context.\\[2pt]

            Output only a single integer from 0 to 5.\\
            /no\_think \\
            %%%%%%%%%%%
                % content end
            \rule{\textwidth}{0.4pt}
            \fontsize{9}{10}\selectfont
            %%%%%%%%
                % output
            \textbf{Response}: 0, 1, 2, 3, 4, or 5
        \end{minipage}
        }
    \end{minipage}

    \caption{A prompt for evaluating the quality of a dialogue model response in a conversational turn-taking scenario.}
    \label{app-fig:dialogue-response-evaluation-prompt-5}
\end{figure}

For that ablation only, the returned score is multiplied by $2/5$ before group normalization so that its magnitude
remains on the historical $0$--$2$ scale.

\paragraph{Full-Duplex-Bench v1 response-quality evaluation.}
Our evaluation pipeline uses the benchmark's interruption evaluator~\citep{lin2025fdbench} with \texttt{gpt-4o-2024-08-06}~\citep{openai2024gpt4o}, temperature $0.2$, and seed
$0$. Agent responses are transcribed with \texttt{nvidia/parakeet-tdt-0.6b-v2} before judging. The evaluator asks the
judge to assess the response after a user interruption on a
$0$--$5$ relevance scale, using the following system prompt:

\begin{figure}[h!]
    \centering
    \begin{minipage}{\textwidth}
    \fbox{
        \begin{minipage}{\textwidth}
            \fontsize{10}{10}\selectfont
            %%%%%%%%%
                % Title
            \textbf{Evaluating AI response after user interruption} \\
            \rule{\textwidth}{0.4pt}
            \fontsize{9}{10}\selectfont
            %%%%%%%%%%
                % content start
            The scenario is that the user and AI are having a spoken conversation. The user speaks first, and the AI begins responding. During the AI's response, the user interrupts the AI.\\
            Your task is to evaluate the quality of the AI's response after the interruption.\\[2pt]

            Use the following rating scale (0--5):\\
            0: The AI's response is totally unrelated to the user's interrupting turn.\\
            1: The AI's response is not related to the user's interrupting turn.\\
            2: The AI's response is slightly related to the user's interrupting turn.\\
            3: The AI's response is related to the user's interrupting turn.\\
            4: The AI's response is highly related to the user's interrupting turn.\\
            5: The AI's response is perfectly related to the user's interrupting turn.\\[2pt]

            First, briefly analyze the user's interrupting turn and the AI's response.\\
            Then provide the rating.\\
            %%%%%%%%%%%
                % content end
            \rule{\textwidth}{0.4pt}
            \fontsize{9}{10}\selectfont
            %%%%%%%%
                % output
            \textbf{Response}:\\
            Analysis: [Your analysis] \\
            I would rate the AI's response as [Rating]
        \end{minipage}
        }
    \end{minipage}

    \caption{A prompt for evaluating how well an AI response addresses a user's interrupting turn.}
    \label{app-fig:interruption-evaluation-prompt}
\end{figure}

The user message supplies three fields: the contextual user turn, the interrupting user turn, and the agent response.
The judge score is computed only for clips on which the benchmark evaluator detects a qualifying agent turn; the
takeover rate is reported alongside it so that conditional quality is not read without response propensity. For the
reported Moshi checkpoints this gives 192 scored responses for \method{} and 191 for GRPO, out of 200 interruption
clips. This evaluator is neither the reward model nor the reward rubric used in training.

\section{Sensitivity of Synthetic Pause Evaluation to the Streaming Protocol}
\label{app:pause}

\begin{tcolorbox}[colback=gray!10, colframe=black!70!white]
Synthetic-pause scores are sensitive to the tested streaming settings. Natural CANDOR clips provide a complementary basis for assessing pause handling.
\end{tcolorbox}

Full-Duplex-Bench v1~\citep{lin2025fdbench} measures pause handling with two subsets. \emph{Candor} pause clips are excerpts from real human conversations (the CANDOR corpus~\citep{reece2023candor}); \emph{synthetic} (Syn) pause clips are constructed from text-to-speech audio with an artificially inserted silence. In our experiments, Syn scores appear sensitive to conversation-start behavior and serving settings. We therefore interpret them alongside the natural-pause results and retain them for comparison with prior reported numbers.

\paragraph{Saturation in the evaluated Moshi runs.} Under the official inference protocol (real-time streaming from $t{=}0$, output capped at input length, no trimming), the base Moshi model scores $0.416$ (compared with the published $0.445$), while the Moshi RL runs in our main comparison and the released Moshi +RL Seamless checkpoint score $1.000$. The latter was reported at $0.307$~\citep{ohashi2026multifacetedinteractivityalignmentfullduplex}. This difference from the reported score, together with the similar scores among these Moshi runs, suggests limited discrimination under our reproduced protocol.

\paragraph{Possible contributions from clip construction.} Two properties of the synthetic clips may contribute to the observed saturation. First, the TTS audio and the mechanically inserted silence may differ from natural hesitations in their prosodic and syntactic continuation cues. Second, the annotated hesitation sits only a second or two into a short clip, so it coincides with the conversation-start boundary. Conversation-start greetings may contribute to the measured takeover rate: with the serving stack's silence warm-up removed, even the \emph{base} model greets and scores $0.985$ (Table~\ref{tab:pause_sensitivity}), and that greeting, over the synthetic clip, is counted as a takeover. These factors may affect the score in addition to the model's handling of a mid-utterance pause.

\paragraph{Comparison across serving settings.} We varied the inference protocol for the released checkpoint and the base model (Table~\ref{tab:pause_sensitivity}): client-side silence of $0$ or $3$\,s, with and without output trimming; and the serving stack's internal silence warm-up removed, default ($1$\,s), or extended to $3$\,s. Among these settings, the base model is closer to its published $0.445$ without greeting exclusion, while the +RL checkpoint is closer to its published $0.307$ with greeting exclusion. This pattern suggests sensitivity to greeting treatment, although it does not establish which settings produced the published results.

\begin{table}[h]
\centering
\caption{Pause Syn TOR for the released checkpoint and base Moshi under the tested protocol variants. ``prime'' = server-side silence warm-up stepped into the model before the conversation; ``greet-excl.'' drops a conversation-start greeting utterance before scoring. Published values are Moshi $0.445$ and +RL $0.307$.}
\label{tab:pause_sensitivity}
\resizebox{\linewidth}{!}{%
\begin{tabular}{lcccccc}
\toprule
Model & official & official + greet-excl. & no-trim, prepend 3s & trim, prepend 3s & official, no prime & official, 3s prime \\
\midrule
Moshi (base) & 0.416 & 0.263 & 0.409 & 0.139 & 0.985 & --- \\
+RL Seamless (released) & 1.000 & 0.292 & 0.993 & 0.496 & 0.993 & 0.978 \\
\bottomrule
\end{tabular}%
}
\end{table}

\paragraph{Natural-pause comparison.} The Candor clips are drawn from the middle of long natural conversations, which reduces the influence of conversation-start greetings and preserves natural continuation cues. This subset shows more variation across the evaluated models, with lower takeover rates for \method{} than for the base and reproduced GRPO models. We therefore emphasize CANDOR pause TOR when discussing pause handling and mark Syn TOR ($^{*}$) in the main table to indicate these interpretation caveats.

\section{Official Turn Metric and Post-Boundary Word Activity}
\label{app:turn}

\begin{tcolorbox}[colback=gray!10, colframe=black!70!white]
Under our reproduced streaming protocol, the official metric can count speech before the user finishes. The post-boundary word-activity diagnostic provides a complementary view of speech after the annotated turn end, but does not identify newly initiated responses.
\end{tcolorbox}

The official Full-Duplex-Bench v1 smooth-turn-taking score~\citep{lin2025fdbench} marks a sample as a successful turn if the agent produces \emph{any} speech span of at least one second (or more than three words) anywhere in the clip, and reports latency as the first word's time minus the user's turn end, clamped at zero. In the Candor clips the user's turn ends roughly halfway through a ten-second excerpt, and early speech is frequent among the evaluated RL-trained models (Table~\ref{tab:turn_strict}). Consequently, these quantities have limited discrimination in our reproduced protocol: the takeover rate mostly reports whether the model speaks at all, and the reported latency is zero or near zero for the models in Table~\ref{tab:fdb1_highlight}.

As a complementary diagnostic, we score post-boundary word activity on the same 119 outputs. We transcribe each generated waveform
with \texttt{nvidia/parakeet-tdt-0.6b-v2} and take the annotated user turn end $\tau$ from the benchmark metadata. Let
$W^+=((a_j,b_j))_{j=1}^{n_+}$ be the ordered suffix of ASR word intervals satisfying $a_j\geq\tau$. We apply the following elapsed-time-or-word-count criterion to that suffix:
\begin{equation}
q_{\mathrm{post}}=\textbf{1}\!\left[n_+>0\ \land\
\left(b_{n_+}-a_1\geq1\,\mathrm{s}\ \lor\ n_+>3\right)\right].
\end{equation}
We report the mean of $q_{\mathrm{post}}$ as a post-boundary
word-activity rate. For qualifying clips, we report the median offset $a_1-\tau$. ``Speaks before turn end'' is measured separately from the first ASR word. This diagnostic identifies qualifying words after the annotated boundary. It does not require a new speech episode to begin after $\tau$. Moreover, $b_{n_+}-a_1$ measures elapsed time and can include silence between words. We therefore do not interpret $q_{\mathrm{post}}$ alone as the rate of newly initiated responses to a completed user turn.

\begin{table}[h]
\centering
\caption{Post-boundary word activity on the same 119 generations. ``Speaks before turn end'' is the fraction of clips whose first ASR word precedes the user's turn end. Post-boundary word activity applies the elapsed-time-or-word-count criterion to words starting after that boundary; it may include continuation of earlier speech. The reported offset is the median time to the first post-boundary word among qualifying clips. Bold and underline mark the best and second-best values within each family for the post-boundary measures.}
\label{tab:turn_strict}
\resizebox{\linewidth}{!}{%
\begin{tabular}{lccc}
\toprule
Model & Official TOR & Speaks before turn end & Post-boundary word activity $\uparrow$ / first-word offset $\downarrow$ \\
\midrule
Moshi & 0.387 & 38.7\% & \phantom{0}9.2\% / \underline{0.63\,s} \\
+ GRPO & 0.992 & 99.2\% & \underline{69.7\%} / 1.27\,s \\
+RL Seamless (released) & 0.992 & 99.2\% & 57.1\% / 1.65\,s \\
\rowcolor{pastelblue}+ \method{} & 1.000 & 100.0\% & \textbf{98.3\%} / \textbf{0.57\,s} \\
\midrule
PersonaPlex & 0.882 & 89.1\% & 34.5\% / 0.85\,s \\
+ GRPO & 0.882 & 89.9\% & \underline{36.1\%} / \underline{0.70\,s} \\
\rowcolor{pastelblue}+ \method{} & 0.714 & 79.8\% & \textbf{47.1\%} / \textbf{0.57\,s} \\
\bottomrule
\end{tabular}}
\end{table}

\section{Evaluation Sample Sizes, Paired Uncertainty, and Training Seeds}
\label{app:statistics}

\paragraph{Benchmark sample sizes.}
The complete Full-Duplex-Bench v1 evaluation contains 137 synthetic-pause clips, 216 CANDOR-pause clips, 55
backchannel clips, 119 smooth-turn-taking clips, and 200 user-interruption clips. All systems in a paired comparison are
evaluated on the same clip identifiers. Interruption response quality is conditional on the benchmark detecting a
qualifying agent turn; this yields 192 judged responses for Moshi + \method{} and 191 for Moshi + GRPO.

\paragraph{Paired clip bootstrap.}
We use paired bootstrap resampling~\citep{koehn2004bootstrap} to estimate uncertainty in metric differences.
For the two headline restraint metrics, we recover the per-clip binary decisions produced by the official evaluators and
resample matched clip identifiers jointly with replacement for 10{,}000 iterations (bootstrap seed $0$). Table~\ref{tab:bootstrap}
reports the observed difference and percentile $95\%$ interval for the fixed 100-epoch Moshi checkpoints. These
intervals quantify benchmark clip-sampling uncertainty for the evaluated checkpoints; they do not represent variation
across independent training runs.

\begin{table}[h]
\centering
\small
\begin{tabular}{lrrrr}
\toprule
Metric & $n$ & \method{} & GRPO & Difference, 95\% CI \\
\midrule
CANDOR pause TOR$\downarrow$ & 216 & 0.306 & 0.454 & $-0.148\;[-0.222,-0.074]$ \\
Backchannel TOR$\downarrow$ & 55 & 0.127 & 0.309 & $-0.182\;[-0.327,-0.036]$ \\
\bottomrule
\end{tabular}
\caption{Paired bootstrap over Full-Duplex-Bench v1 clips for Moshi. A negative difference favors \method{} for both
takeover-rate metrics.}
\label{tab:bootstrap}
\end{table}

\paragraph{Training-seed replication.}
We independently train the reported event-causal objective, the fixed-window routing ablation, and GRPO with
seeds $42$, $43$, and $44$. Each method is repeated with its own training configuration and evaluated at 100 epochs on
Full-Duplex-Bench v1. GRPO reproduces the training method of \citet{ohashi2026multifacetedinteractivityalignmentfullduplex},
using the same pretrained backbone as \method{}. Response quality is rescored uniformly with \texttt{gpt-4o-2024-08-06}, temperature $0.2$,
and evaluation seed $0$. Table~\ref{tab:seed-replication} reports the mean and sample standard deviation across the
three training seeds.

\begin{table*}[h]
\centering
\small
\resizebox{\linewidth}{!}{%
\begin{tabular}{lrrrrrrr}
\toprule
Credit rule & Pause TOR$\downarrow$ & BC TOR$\downarrow$ & BC freq.$\uparrow$ & Turn TOR$\uparrow$ & Interrupt TOR$\uparrow$ & Judge$\uparrow$ & Int. lat.$\downarrow$ \\
\midrule
GRPO & $0.495\!\pm\!0.050$ & $0.297\!\pm\!0.011$ & $0.131\!\pm\!0.020$ & $0.992\!\pm\!0.000$ & $0.985\!\pm\!0.026$ & $3.927\!\pm\!0.033$ & $0.631\!\pm\!0.026$ \\
fixed window & $0.500\!\pm\!0.060$ & $0.261\!\pm\!0.107$ & $0.132\!\pm\!0.014$ & $0.958\!\pm\!0.030$ & \textbf{$\textbf{0.990}\!\pm\!0.010$} & \textbf{$\textbf{4.300}\!\pm\!0.024$} & $0.480\!\pm\!0.071$ \\
\rowcolor{pastelblue}event-causal (\method{}) & \textbf{$\textbf{0.329}\!\pm\!0.083$} & \textbf{$\textbf{0.109}\!\pm\!0.018$} & \textbf{$\textbf{0.145}\!\pm\!0.020$} & \textbf{$\textbf{1.000}\!\pm\!0.000$} & $0.987\!\pm\!0.023$ & $3.914\!\pm\!0.260$ & \textbf{$\textbf{0.466}\!\pm\!0.061$} \\
\bottomrule
\end{tabular}%
}
\caption{Moshi replication over training seeds $42$, $43$, and $44$. Entries are mean $\pm$ sample standard
deviation. Judge denotes the gpt-4o-2024-08-06 response-quality score. Int. lat. denotes post-interruption response latency. Bold marks the best mean in each column.}
\label{tab:seed-replication}
\end{table*}

The restraint improvement is repeated in every seed. Event-causal routing has lower CANDOR pause TOR and lower
backchannel TOR than GRPO for each of the three paired seeds. It also preserves a perfect smooth-turn response
rate and reduces mean post-interruption response latency. Its mean judged quality is close to flat GRPO, with greater seed
variation. Fixed-window routing attains the highest conditional judged quality but gives up pause restraint,
backchannel restraint, and turn response. Thus the replication supports the multi-axis timing result, while not
claiming that event-causal routing maximizes every metric.

\section{Sensitivity to the GRPO Learning Rate}
\label{app:lrsweep}

\begin{tcolorbox}[colback=gray!10, colframe=black!70!white]
Across the tested learning rates, GRPO shows different trade-offs across metrics. The reported \method{} configuration retains advantages on several measures.
\end{tcolorbox}

\S\ref{sec:objective} uses separately tuned learning rates for the two methods. To assess sensitivity to this choice,
Table~\ref{tab:lrsweep} compares four GRPO learning rates, including the rate used by \method{}, while holding
the other GRPO settings fixed.

% Auto-generated by scripts/make_lrsweep_table.py
\begin{table}[ht]
\centering
\small
\begin{tabular}{ll cccccc}
\toprule
Method & $\eta$ & Pause$\downarrow$ & BC TOR$\downarrow$ & BC freq$\uparrow$ & Turn$\uparrow$ & Judge$\uparrow$ & Int. lat.$\downarrow$ \\
\midrule
GRPO & $1\times10^{-6}$ & 0.454 & 0.309 & 0.110 & \underline{0.992} & \underline{3.942} & 0.660 \\
GRPO & $5\times10^{-7}$ & 0.574 & 0.400 & 0.166 & 0.924 & 3.894 & 1.067 \\
GRPO & $2\times10^{-6}$ & 0.528 & \underline{0.073} & \textbf{0.174} & \underline{0.992} & 3.872 & 0.491 \\
GRPO & $4\times10^{-6}$ & \underline{0.431} & \textbf{0.000} & \underline{0.171} & \underline{0.992} & 3.890 & \underline{0.464} \\
\midrule
\rowcolor{pastelblue}\method{} & $4\times10^{-6}$ & \textbf{0.306} & 0.127 & 0.134 & \textbf{1.000} & \textbf{4.083} & \textbf{0.439} \\
\bottomrule
\end{tabular}
\caption{The GRPO baseline swept over learning rates, including the rate \method{} uses. Everything except the step size is held fixed. Int. lat. denotes post-interruption response latency. Best values are bold and second-best values are underlined; ties share the same marking.}
\label{tab:lrsweep}
\end{table}

Reducing the GRPO learning rate from $10^{-6}$ to $5\times10^{-7}$ worsens most reported metrics, although
backchannel frequency increases. At $4\times10^{-6}$, GRPO has a lower backchannel takeover rate than
\method{} ($0.000$ versus $0.127$) and a higher backchannel frequency. At that rate, however, its pause
TOR remains higher ($0.431$ versus $0.306$), its turn-response rate is slightly lower, and its judged quality
is lower ($3.890$ versus $4.083$). These results suggest that the comparison depends on the metric, while
supporting the observed advantages of \method{} across several measures within the tested range.

\section{Mechanism Diagnostics: Where Each Objective Moves the Policy}
\label{app:diagnostics}

\begin{tcolorbox}[colback=gray!10, colframe=black!70!white]
In the reported runs, control KL is larger than the probability-weighted content KL contribution, and semantic feedback is associated with lower repetition. These observations are consistent with differentiated updates to the two factors under the tested settings.
\end{tcolorbox}

The main results show that \method{} improves timing without losing content, but they do not by themselves show
\emph{why}. Table~\ref{tab:diagnostics} looks inside training. All quantities are averaged over the optimizer updates of
the last $30\%$ of each run, except repetition, which is measured on the text the model actually produced during the
benchmark's interruption task.

For GRPO, we report the total token-policy KL recorded in these runs. For \method{}, we report
control KL and the conditional content KL weighted by the current policy's probability of selecting content. The same probability factorization could also be applied
to a GRPO checkpoint for separate diagnostics, but those measurements are not reported here.
For \method{}, the recorded control KL exceeds the weighted content contribution. The two factors carry distinct regularization terms and use different policy-loss eligibility rules, matching the intended control/content split.

Each update touches a few dozen control decisions and under ten content decisions per rollout.
Across the reported runs, event-causal updates select approximately $20$--$30\%$ of the frames
used by a flat update. These differences motivate tuning learning rates separately (\S\ref{sec:objective}),
although the number of selected frames alone does not determine policy movement.

The repetition column provides another view of the semantic-reward comparison. Trained on timing alone, \method{} repeats trigrams an order of magnitude more often than any other system here, which is the degradation the judged-quality column of Table~\ref{tab:rllm_abl} reports, seen from the text side. At the reported weights, routing the judge reward to the content factor brings repetition below even the flat baselines while the control-side KL stays essentially unchanged. This pattern is consistent with the intended role of level-specific routing. It is not a guarantee that the behaviors are independent because both factor losses update the same backbone. Table~\ref{tab:semmax} stress-tests that limitation by deliberately making the content objective dominant.

\begin{table}[h]
\centering
\small
\begin{tabular}{lccccc}
\toprule
 & \multicolumn{2}{c}{Reference KL} & \multicolumn{2}{c}{Decisions per update} & \\
\cmidrule(lr){2-3}\cmidrule(lr){4-5}
System & control & content contrib. & control & content & Repetition-3 $\downarrow$ \\
\midrule
GRPO w/o $R_{llm}$ & \multicolumn{2}{c}{$2.6\times10^{-2}$ (single policy)} & \multicolumn{2}{c}{$\sim$200 (all frames)} & \underline{0.007} \\
GRPO & \multicolumn{2}{c}{$2.1\times10^{-2}$ (single policy)} & \multicolumn{2}{c}{$\sim$200 (all frames)} & 0.009 \\
\midrule
\method{} w/o $R_{llm}$ & $9.8\times10^{-2}$ & $1.0\times10^{-2}$ & 38.7 & 9.1 & 0.098 \\
\rowcolor{pastelblue}\method{} & $8.6\times10^{-2}$ & $1.4\times10^{-2}$ & 40.6 & 8.6 & \textbf{0.002} \\
\bottomrule
\end{tabular}
\caption{What each objective actually moves. Reference KL reports control KL and the content KL contribution weighted by $\pi_\theta^{\mathrm{ctrl}}(\con)$. KL values and decision counts are averaged over the optimizer updates of
the final $30\%$ of training. Repetition is the fraction of trigrams that recur within a response on the
Full-Duplex-Bench v1 interruption task. The flat-baseline rows report total token-policy KL. Separate control and content KL were not recorded for these rows. The best repetition value is bold and the second-best
is underlined. These diagnostics describe the reported, balanced reward setting rather than invariance to reward scaling.}
\label{tab:diagnostics}
\end{table}

\paragraph{Sensitivity to the behavioral prior.}
At the $4\times10^{-6}$ rate tuned on Moshi, PersonaPlex collapses after roughly 1.75k optimizer steps into continuous
audio with an empty text stream. Suppression improves while initiation decays, so the policy buys reward by speaking
less. Halving the rate removes the abrupt transition but converges toward the same near-silent endpoint. Rebalancing
toward initiation keeps both components healthy. Since the models share the backbone but PersonaPlex adds a more
talkative, floor-holding conversational prompt, that behavioral prior may contribute to the narrower stability margin:
the early suppression correction can overshoot from avoiding overlap to producing no text. The endpoint benchmark
scores in Figure~\ref{fig:dynamics} distinguish this failure from useful restraint.

\section{What Interaction Looks Like in the Training Data}
\label{app:datastats}

\begin{tcolorbox}[colback=gray!10, colframe=black!70!white]
Timing distributions vary across the examined corpora, suggesting that a single target value may capture only part of conversational behavior.
\end{tcolorbox}

The rewards in \S\ref{sec:eventcausal} assume that quantities like ``a well-timed response'' or ``a brief
acknowledgment'' have a stable meaning. Figure~\ref{fig:datastats} checks that assumption against the corpora
themselves, measuring the same five quantities the benchmark scores on the two Seamless subsets we train on and on
Fisher English~\citep{cieri2004fisher}. For Seamless we use the released per-channel voice-activity annotations; Fisher ships utterance-level
timestamps with speaker labels, so we read those directly.

Three observations matter for how the results should be read. First, the distributions are wide: a single target value
for ``correct'' timing describes the centre of a broad distribution rather than a property every conversation shares.
Second, corpora differ from each other in exactly the dimensions we optimize. Fisher hands the floor over at a median
offset of essentially zero, with slightly more than half of its transitions overlapping, whereas the Seamless subsets
leave a positive gap of a third to two thirds of a second and overlap on roughly a third of transitions. Fisher speakers
also hold the floor far longer after being interrupted, a median of about five and a half seconds against three in
Seamless. Telephone conversation between strangers and prompted face-to-face interaction are simply different interaction
regimes.
Third, the improvised and naturalistic Seamless subsets differ from each other as well, with roughly twice the
backchannel rate in the improvised half. Interaction style is a property of the setting, not only of the speakers, which
is the empirical version of the point we raise in \S\ref{sec:discussion_pref}.

The backchannel-length panel describes the short episodes identified under our one-second criterion.
These distributions provide context for the duration-based reward, although they do not independently establish
that one second is the appropriate boundary between an acknowledgment and an attempt to take the floor.

Table~\ref{tab:datastats} reports the corresponding distribution statistics for all three corpora.
\begin{table}[h]
\centering
\small
\resizebox{\linewidth}{!}{%
\begin{tabular}{llccc}
\toprule
Quantity & Statistic & Seamless (improvised) & Seamless (naturalistic) & Fisher \\
\midrule
Own pause (s) & mean $\pm$ sd & 1.07 $\pm$ 1.02 & 0.85 $\pm$ 0.81 & 1.40 $\pm$ 1.22 \\
 & median (IQR) & 0.64 (0.39--1.35) & 0.55 (0.36--0.96) & 0.92 (0.46--1.99) \\
 & p10 / p90 & 0.26 / 2.60 & 0.26 / 1.80 & 0.29 / 3.45 \\
\addlinespace
Backchannels per minute & mean $\pm$ sd & 4.54 $\pm$ 2.24 & 2.82 $\pm$ 2.11 & 3.86 $\pm$ 1.29 \\
 & median (IQR) & 4.21 (2.77--6.35) & 2.38 (1.28--3.91) & 3.80 (3.00--4.68) \\
 & p10 / p90 & 1.82 / 7.51 & 0.43 / 6.08 & 2.30 / 5.41 \\
\addlinespace
Backchannel length (s) & mean $\pm$ sd & 0.47 $\pm$ 0.25 & 0.49 $\pm$ 0.22 & 0.54 $\pm$ 0.30 \\
 & median (IQR) & 0.44 (0.32--0.64) & 0.48 (0.35--0.64) & 0.56 (0.28--0.84) \\
 & p10 / p90 & 0.12 / 0.83 & 0.19 / 0.83 & 0.12 / 0.94 \\
\addlinespace
Handover offset (s) & mean $\pm$ sd & 0.81 $\pm$ 2.00 & 1.16 $\pm$ 2.10 & 0.50 $\pm$ 2.05 \\
 & median (IQR) & 0.32 (-0.35--1.83) & 0.68 (-0.19--2.66) & -0.05 (-0.63--0.92) \\
 & p10 / p90 & -1.44 / 4.07 & -1.34 / 4.42 & -1.66 / 4.10 \\
\addlinespace
Time until yielding (s) & mean $\pm$ sd & 5.45 $\pm$ 7.01 & 7.92 $\pm$ 9.94 & 11.20 $\pm$ 15.14 \\
 & median (IQR) & 2.88 (1.60--6.59) & 3.58 (1.74--10.01) & 5.67 (2.43--13.69) \\
 & p10 / p90 & 1.21 / 12.02 & 1.24 / 20.71 & 1.44 / 26.90 \\
\addlinespace
\midrule
Overlapping handovers & fraction & 0.36 & 0.29 & 0.52 \\
Conversations & count & 146 & 135 & 150 \\
\bottomrule
\end{tabular}}
\caption{Distribution descriptors for the quantities in Figure~\ref{fig:datastats}. Spread matters as much as location here: the standard deviations are comparable to the means, so no single value characterises ``correct'' timing in any of the corpora. The overlap row is the fraction of floor transfers where the next speaker starts before the current one stops.}
\label{tab:datastats}
\end{table}

\paragraph{Construction of the Wasserstein diagnostic.}
Table~\ref{tab:distmatch} is a complementary distributional diagnostic rather than a replacement for the benchmark's
official task metrics. With random seed $0$, we sample at most 150 conversations from each of Seamless naturalistic,
Seamless improvised, Fisher part 1, and Fisher part 2, then pool the resulting empirical marginals. After filtering,
the reference contains 17{,}439 floor-transfer offsets, 14{,}058 interruption-overlap durations, 573 conversation-level
backchannel rates, 13{,}669 short-episode durations, and 58{,}223 pause events. Human and model speech spans are merged
across gaps of at most 1\,s. Model-side distributions are computed from the already generated benchmark outputs: 119
turn clips, all 55 backchannel clips, and 277 annotated pause events per system; the backchannel-length marginal contains
46--99 short episodes across the three Moshi systems and 61--67 across the PersonaPlex systems. A sustained turn must
last over 1\,s. A turn clip with no qualifying response is assigned the fixed 6\,s ceiling so that silence cannot reduce
the distance by contributing no sample; using a 4\,s or 8\,s ceiling changes the absolute distances but preserves the
within-family ordering. A backchannel clip with no generated speech contributes a zero event-per-minute rate rather
than being dropped. Turn and backchannel-duration distances are in
seconds, backchannel rate is in events per minute, and pause intrusion is a probability-mass distance. We omit the
interruption-yield marginal because none of the evaluated outputs was still holding the floor at the barge-in instant,
leaving no model-side yield-duration samples. The result measures agreement with these pooled one-dimensional
marginals; it does not claim to measure the full joint or context-conditioned distribution of human interaction.

\paragraph{Reading the model distances.}
On Moshi, the base model is far from the human turn-timing marginal ($4.70$) because it often does not answer; GRPO
closes most of that gap ($1.78$), and \method{} closes it further ($0.90$). Its backchannel rate is also more than four
times closer to the human reference than either alternative. GRPO's smaller pause-intrusion distance has a less
desirable cause: it almost never speaks during a hesitation, while the human intrusion rate is small but nonzero. On
PersonaPlex, \method{} is closer on turn timing and backchannel rate, flat GRPO on backchannel length, and the base model
on pause intrusion. These are marginal comparisons and should be read with the task metrics rather than as a single
aggregate measure of conversational naturalness.

\section{How Coarse Should the Judge's Scale Be?}
\label{app:rubric}

\begin{tcolorbox}[colback=gray!10, colframe=black!70!white]
In the Moshi rubric comparison, the $0$--$2$ scale gives a more favorable balance across the reported metrics.
\end{tcolorbox}

Table~\ref{tab:rllm_scale} asks how coarse the judge's scale should be on Moshi. The $0$--$2$ run shows lower pause and backchannel takeover rates, while the $0$--$5$ run has a slightly lower backchannel JSD. We therefore use
$0$--$2$ for both reported models. The $0$--$5$ result is a Moshi rubric ablation only and is not the reported
PersonaPlex configuration.

% Rubric range for the LLM judge: 0-2 vs 0-5, HiPLEX only
\begin{table*}[t]
\centering
\small
\resizebox{\linewidth}{!}{%
\begin{tabular}{l cc ccc cc ccc}
\toprule
 & \multicolumn{2}{c}{Pause} & \multicolumn{3}{c}{Backchannel} & \multicolumn{2}{c}{Smooth Turn Taking} & \multicolumn{3}{c}{User Interruption} \\
\cmidrule(lr){2-3}\cmidrule(lr){4-6}\cmidrule(lr){7-8}\cmidrule(lr){9-11}
Judge rubric & Syn TOR$\downarrow$ & Candor TOR$\downarrow$ & TOR$\downarrow$ & Freq$\uparrow$ & JSD$\downarrow$ & TOR$\uparrow$ & Latency$\downarrow$ & TOR$\uparrow$ & Judge$\uparrow$ & Latency$\downarrow$ \\
\midrule
\rowcolor{pastelblue}\method{} with $0$--$2$ & \textbf{1.000} & \textbf{0.306} & \textbf{0.127} & \textbf{0.134} & \underline{0.734} & \textbf{1.000} & \textbf{0.000} & \textbf{0.960} & \textbf{4.083} & \textbf{0.439} \\
\method{} with $0$--$5$ & \textbf{1.000} & \underline{0.407} & \underline{0.200} & \underline{0.131} & \textbf{0.733} & \textbf{1.000} & \textbf{0.000} & \underline{0.955} & \underline{4.021} & \underline{0.445} \\
\bottomrule
\end{tabular}%
}
\caption{Effect of the judge rubric range on \method{}. The $0$--$2$ rubric performs better on most reported metrics, with ties on several metrics and a slightly higher backchannel JSD. This comparison does not establish why the rubrics differ. Changes in rating patterns may affect the training signal, but a smaller raw-score spread alone does not imply smaller standardized advantages. Best values are bold and second-best values are underlined; ties share the same marking.}
\label{tab:rllm_scale}
\end{table*}

\section{Sensitivity to Stronger Semantic Feedback}
\label{app:semmax}

\begin{tcolorbox}[colback=gray!10, colframe=black!70!white]
In the tested PersonaPlex variants, stronger semantic feedback sometimes raises judged quality alongside lower response rates. This suggests sensitivity to the balance between content and timing objectives.
\end{tcolorbox}

The routed semantic reward raises judged quality (\S\ref{sec:results_main}), which invites an obvious question: how far
does it go if we simply weight it more? We examine PersonaPlex variants with stronger semantic advantages ($\times1.5$ or $\times2$),
increased content-loss weighting, or weaker content KL regularization. We also evaluate continued training
from the reported checkpoint. Whereas Table~\ref{tab:diagnostics} characterizes the balanced setting reported in the main results,
Table~\ref{tab:semmax} is a stress test outside that operating regime.

One tested checkpoint reaches a judged-quality score of $4.488$, close to the value reported by
\citet{ohashi2026multifacetedinteractivityalignmentfullduplex} for their PersonaPlex system. At the same checkpoint,
the model takes the floor in $58.3\%$ of backchannel clips, compared with $1.8\%$ for our reported configuration.
Its turn-response rate is $46.7\%$, decreasing to $3.3\%$ ten epochs later. Several later checkpoints also
show low response rates, suggesting that the quality gains can coincide with reduced participation.

These results motivate reading conditional judged quality alongside timing and response-rate metrics.
They also suggest sensitivity to the relative strength of the content objective. Since the policy factors share
parameters, reward weighting, loss weighting, and KL regularization may influence both content and timing.
The tested variants illustrate this trade-off without establishing a general limit on semantic feedback.

\begin{table}[h]
\centering
\small
\resizebox{\linewidth}{!}{%
\begin{tabular}{lccccc}
\toprule
Variant (checkpoint) & Pause Candor TOR$\downarrow$ & BC TOR$\downarrow$ & BC Freq$\uparrow$ & Turn TOR$\uparrow$ & Judge$\uparrow$ \\
\midrule
semantic $\times1.5$, loose content KL (e20) & 0.389 & 0.583 & \underline{0.071} & \underline{0.467} & \textbf{4.488} \\
semantic $\times1.5$, loose content KL (e30) & 0.167 & \textbf{0.000} & 0.013 & 0.067 & 3.955 \\
semantic $\times1.5$, loose content KL (e90) & 0.088 & \textbf{0.000} & 0.046 & 0.017 & 3.500 \\
semantic $\times2$, content weight $1.5$ (e90) & 0.181 & \textbf{0.000} & 0.062 & 0.076 & --- \\
semantic $\times2$, content weight $1.5$ (e100) & 0.199 & \underline{0.036} & 0.059 & 0.017 & \underline{4.320} \\
continued from e100 of the main run (e130) & \textbf{0.074} & \textbf{0.000} & 0.011 & 0.033 & 3.458 \\
continued from e100 of the main run (e150) & \underline{0.185} & \textbf{0.000} & 0.016 & 0.033 & 2.905 \\
\midrule
\rowcolor{pastelblue}\method{} as reported in Table~\ref{tab:fdb1_highlight} (e100) & 0.218 & \underline{0.018} & \textbf{0.106} & \textbf{0.714} & 2.938 \\
\bottomrule
\end{tabular}%
}
\caption{PersonaPlex sensitivity to stronger semantic feedback and continued training. A tested checkpoint reaches
judged quality $4.488$ alongside a higher backchannel takeover rate, while several later checkpoints have low
turn-response rates. These results complement the diagnostics in Table~\ref{tab:diagnostics} by examining
settings outside the reported configuration. The final blue row is the configuration we report in the main table. Best values are bold and second-best values are
underlined; ties share the same marking. Rows are scored on a $25\%$ sample of the benchmark, which is why one judge
entry is missing.}
\label{tab:semmax}
\end{table}

\section{Discussion}

\paragraph{Toward models that fit the listener, not the average listener.}
\label{sec:discussion_pref}
There is a further question that our results raise but do not answer. The benchmarks used here score fixed interaction
outcomes, while our distributional analysis compares model timing with pooled human marginals. Neither captures how
preferred timing varies across people and situations. Some people want to be heard out, and a partner who acknowledges
constantly may seem intrusive. Others enjoy a fast, talkative exchange in which frequent overlap signals engagement.
The same person shifts between these modes with the situation. A learner practising a language may want to be corrected
as soon as a phrase comes out wrong, which in another context would feel intrusive. Human timing varies with the
speaker and situation, and a conversational model should eventually adapt to that variation.

This suggests two directions. First, interaction preferences could condition the model's behavior through explicit
settings for backchannel frequency and readiness to take the floor, or through spoken requests such as ``let me finish
my thoughts'' and ``feel free to jump in''. Our factorization exposes emission control as a probability factor that
could be conditioned on these preferences. Because control and content share parameters, whether such conditioning
preserves response quality remains to be tested. Second, evaluation needs to test whether a model can \emph{move}
between interaction profiles on request. This requires data with varied preferences and metrics that measure behavior
under each requested profile.

\end{document}